\documentclass[conference]{IEEEtran}
\IEEEoverridecommandlockouts
\usepackage[english]{babel}
\usepackage[utf8]{inputenc}
\usepackage{cite}
\usepackage{graphicx}
\usepackage{textcomp}
\usepackage{xcolor}
\usepackage[justification=centering]{caption}
\usepackage{subcaption}
\usepackage{array}

\usepackage[normalem]{ulem} % \sout (strikethrough text)
\usepackage{fancyhdr}

\usepackage{algorithm, algorithmic}
\usepackage{amsmath,amsthm,amssymb,amsfonts}
\usepackage{mathtools}
\usepackage{braket}

\usepackage[colorlinks=true,allcolors=blue]{hyperref}

\definecolor{darkgreen}{RGB}{43, 148, 136}
\definecolor{pink}{RGB}{186, 76, 129}
\definecolor{lightgreen}{RGB}{78, 186, 114}
\definecolor{myyellow}{RGB}{180, 180, 0}
\definecolor{darkred}{RGB}{180, 0, 0}
\definecolor{darkblue}{RGB}{0, 0, 180}

\newcommand{\C}{\mathbb{C}}

\newcommand{\tn}[1]{\boldsymbol{\mathcal{#1}}}

\newcommand{\tsr}[1]{\boldsymbol{#1}}

\newcommand{\mat}[1]{\boldsymbol{#1}}
\newcommand{\vcr}[1]{\boldsymbol{#1}}
\makeatletter
\newsavebox{\@brx}
\newcommand{\llangle}[1][]{\savebox{\@brx}{\(\m@th{#1\langle}\)}%
  \mathopen{\copy\@brx\kern-0.5\wd\@brx\usebox{\@brx}}}
\newcommand{\rrangle}[1][]{\savebox{\@brx}{\(\m@th{#1\rangle}\)}%
  \mathclose{\copy\@brx\kern-0.5\wd\@brx\usebox{\@brx}}}
\makeatother

\usepackage{listings}
\definecolor{mygreen}{rgb}{0,0.2,0}
\definecolor{mygray}{rgb}{0.5,0.5,0.5}
\definecolor{mymauve}{rgb}{0.58,0,0.82}
\definecolor{mypurple}{rgb}{0.38,0,0.32}
\definecolor{myblue}{rgb}{0.1,0,0.32}
\newcommand{\costyle}{\footnotesize\ttfamily\bfseries}
\newcommand{\kwstyle}{\costyle\textcolor{myblue}}

\def\BibTeX{{\rm B\kern-.05em{\sc i\kern-.025em b}\kern-.08em
    T\kern-.1667em\lower.7ex\hbox{E}\kern-.125emX}}

\begin{document}

\title{Efficient 2D Tensor Network Simulation of Quantum Systems
% TODO
% \thanks{Identify applicable funding agency here. If none, delete this.}
}

\author{
\IEEEauthorblockN{
    Yuchen Pang\IEEEauthorrefmark{1},
    Tianyi Hao\IEEEauthorrefmark{1},
    Annika Dugad\IEEEauthorrefmark{1},
    Yiqing Zhou\IEEEauthorrefmark{1}
    and
    Edgar Solomonik\IEEEauthorrefmark{1}
}
\IEEEauthorblockA{\IEEEauthorrefmark{1}
    Department of Computer Science,
    University of Illinois at Urbana-Champaign,
    Urbana, IL 61801, USA\\
    Email:
    \{yuchenp2, tianyih2, dugad2, yiqing2, solomon2\}@illinois.edu
}
}

\maketitle

% Copyright
\thispagestyle{fancy}
\lhead{}
\rhead{}
\chead{}
\lfoot{\footnotesize{
SC20, November 9-19, 2020, Is Everywhere We Are
\newline 978-1-7281-9998-6/20/\$31.00 \copyright 2020 IEEE}}
\rfoot{}
\cfoot{}
\renewcommand{\headrulewidth}{0pt}
\renewcommand{\footrulewidth}{0pt}

\begin{abstract}
Simulation of quantum systems is challenging due to the exponential size
of the state space. Tensor networks provide a systematically improvable
approximation for quantum states.
2D tensor networks such as Projected Entangled Pair States (PEPS)
are well-suited for key classes of physical systems and quantum circuits.
However, direct contraction of PEPS networks has exponential cost, while
approximate algorithms require computations with large tensors.
We propose new scalable algorithms and software abstractions for PEPS-based
methods, accelerating the bottleneck operation of contraction and refactorization
of a tensor subnetwork. We employ randomized SVD
with an implicit matrix to reduce cost and memory footprint asymptotically.
Further, we develop a distributed-memory PEPS library and
study accuracy and efficiency of alternative algorithms for PEPS contraction
and evolution on the Stampede2 supercomputer. We also simulate a popular
near-term quantum algorithm, the Variational Quantum Eigensolver (VQE), and
benchmark Imaginary Time Evolution (ITE), which compute ground states of Hamiltonians.
\end{abstract}

\begin{IEEEkeywords}
Tensor network, Projected Entangled Pair States, Quantum system
\end{IEEEkeywords}

\section{Introduction} \label{sction:introduction}

The degrees of freedom required to directly represent a quantum state grow exponentially with respect to the size of the system. However,
for slightly entangled quantum states, polynomial-size representations are possible by using tensor networks~\cite{orus2014tnreview,vidal2003efficient}. 1D tensor networks, also known as \emph{matrix product states} (MPS), have proven successful in modeling 1D strongly correlated quantum systems~\cite{schollwock2011dmrg}, and various types of tensor networks have been invented to generalize this method to higher dimensions~\cite{verstraete2004peps,verstraete2004mpo,vidal2008mera,shi2006tree}. Among these variations, the 2D tensor lattice networks, known as \emph{projected entangled pair states} (PEPS), enable effective representation of many physical systems of interest due to their 2D structure and are therefore widely used in condensed matter physics~\cite{verstraete2004peps,murg2007peps,jordan2008peps}. For the same reason, PEPS are also suitable to model near-term quantum computers with 2D architectures~\cite{preskill2018quantum,shafaei2014qubit}. 
In fact, this method has been recently applied to a large-scale exact simulation of 2D random quantum circuits~\cite{guo2019simulation}.

Computations with PEPS involve high-order tensors and are computationally demanding. For example, exact contraction of PEPS to a scalar to calculate an expectation value is a \#P-complete problem\cite{schuch2007complexity}. Approximations must be taken to avoid exponential cost, but the order of complexity is still high.
Given a PEPS network of bond dimension $r$, i.e., composed of tensors that are of dimension $r\times r\times r\times r$, and maintaining approximate intermediates that have bond dimensions no larger than some $m \propto r^2$, state-of-the-art PEPS contraction algorithms have a cost that scales as $O(m^3r^2)$ via an iterative optimization procedure.

In search for an efficient parallel implementation for these PEPS algorithms, we propose an abstraction called \texttt{einsumsvd} for operations that contract a set of tensors into one and then refactorize it into two new tensors as described in Section~\ref{section:background}. This abstraction encapsulates the most costly operations in many tensor network applications, and it can be implemented with a variety of algorithms. In Section~\ref{section:algorithms}, we discuss an approach for performing \texttt{einsumsvd} that leverages randomized \emph{singular value decomposition} (SVD) with implicit applications of the tensor operator. When applying this idea to the contraction of PEPS, we obtain a new algorithm that minimizes the cost relative to state-of-the-art approaches, while requiring only a single pass over the tensors.
In the same section, we also present an intermediate caching strategy that further lowers the cost of calculating expectation values by trading space for time.

We implement these algorithms as part of a new Python library called Koala for PEPS-based simulations of quantum systems (Section~\ref{section:implementation}). This library supports various PEPS evolution and contraction algorithms, including our improvements, in the sequential/threaded setting using NumPy~\cite{oliphant2006numpy}, the GPU setting using CuPy~\cite{nishino2017cupy}, and the distributed-memory setting using Cyclops~\cite{solomonik2014massively,zhang2019enabling}.
To improve the performance with Cyclops, we propose a parallel approach for tensor orthogonalization that avoids matricization of a high-order tensor, reducing synchronization between processors.
This approach forms small Gram matrices in local memory so that matricization can be done locally while large-tensor contractions are kept distributed.

We evaluate the proposed algorithms and software in Section~\ref{section:results} via a performance study of the PEPS evolution and contraction methods, using single-node and many-node execution on the Stampede2 supercomputer. Our experimental results demonstrate that the Gram matrix method improves the parallel performance of PEPS evolution up to 3.6X, and the randomized SVD approach improves the performance of PEPS contraction up to 22.4X in the sequential/threaded setting and up to 21.3X in the distributed-memory setting (combined with the Gram matrix method). The performance improvement due to the proposed caching optimization is also quantified in this section. For parallel scaling, we perform strong and weak scaling tests with up to 256 nodes with 64 cores per node and show good weak scaling performance with respect to the tensor size (i.e. bond dimension).

As driver applications, we consider a common method for ground state calculation using tensor networks, \emph{imaginary time evolution} (ITE), as well as a simulation of a prominent candidate algorithm for near-term quantum computers, the \emph{variational quantum eigensolver} (VQE). These two applications, which are representative of many algorithms in the domain, are both implemented by the PEPS evolution and contraction primitives provided by our library. In Section \ref{subsec:apps}, we study the numerical accuracy achieved by PEPS simulations with various bond dimensions for ITE and VQE. Our results demonstrate that higher accuracy is achieved by the use of larger tensors (i.e. larger bond dimension) with PEPS.

Overall, this paper makes the following contributions:
\begin{itemize}
    \item a new approach for PEPS contraction that minimizes cost and does not require successive rounds of optimization,
    \item an automatic mechanism to cache intermediate contraction results during PEPS contraction,
    \item novel distributed-memory implementations of multiple PEPS contraction algorithms,
%    for PEPS contraction a distributed parallel approach to PEPS algorithms, enabled by improvements to the costs of PEPS contraction algorithms and a caching strategy for expectation value calculations with PEPS (Section~\ref{section:algorithms});
    \item a software library, Koala, providing sequential and parallel primitives for PEPS-based approximate simulations of quantum systems as well as encapsulating above algorithms via new software abstractions,% (Section~\ref{section:implementation});
    \item to the best of our knowledge, the first study of accuracy and scalability of massively-parallel approximate PEPS algorithms with distributed-memory tensors.
\end{itemize}
A comparison between this work and related work is presented in Section~\ref{section:related-work}.

\section{Background} \label{section:background}

In this section, we provide a brief review on tensor networks and their applications with a focus on the concepts used in this paper. Broader surveys on these topics are available~\cite{orus2014tnreview}.
We list notational conventions employed in the paper in Table~\ref{tab:notation}.
\begin{table}[htbp]
    \centering
    \begin{tabular}{| m{15em} | m{12.5em} |}
        \hline
    	\textbf{Notation} & \textbf{Meaning} \\
    	\hline
    	$\tn{A}$ (bold, calligraphic capital letters) & Tensor network \\
    	\hline 
    	$\tsr{A}$ (bold capital letters) & Matrix or high-order tensor \\
    	\hline
    	$\vcr{a}$ (bold lowercase letters) & Vector \\
    	\hline
    	$a_{ijk}$ & Element of the tensor $\tsr{A}$ \\
    	\hline
    	$a^*$, $\tsr{A}^*$ & Complex conjugate of a scalar $a$ or conjugate transpose of an operator $\tsr{A}$ \\
    	\hline
    	$\tsr{A}\in\C^{d_1\times\cdots\times d_n}$ & Complex tensor $\tsr{A}$ of dimension $d_1\times\cdots\times d_n$ \\
    	\hline
    	$\tsr{A}:\C^{p_1\times\cdots\times p_s}\mapsto\C^{q_1\times\cdots\times q_t}$ & Complex tensor $\tsr{A}$ treated as a linear operator \\
    	\hline
    	$\ket{\psi}$ & Quantum state labeled by $\psi$ \\
    	\hline
        $\bra{\psi}$ & Conjugate transpose of $\ket{\psi}$ \\
        \hline
    \end{tabular}
    \caption{Common notation used in this paper.}
    \label{tab:notation}
\end{table}

\subsection{Quantum States as Tensors}\label{subsec:qsat}

The \emph{state} $\ket{\psi}$ of a quantum computer with $n$ qubits can be described by a unit vector in $\mathbb{C}^{2^n}$. By choosing $2^n$ orthonormal basis vectors/states to be denoted as $\ket{\vcr{i}}$ with $\vcr{i}=i_1\cdots i_n\in\{0,1\}^n$, $\ket{\psi}$ can be written as
\[\ket{\psi} = \sum_{\vcr{i}\in\{0,1\}^n} t^{(\psi)}_{\vcr{i}}\ket{\vcr{i}}.\]
Here, the amplitudes $t^{(\psi)}_{\vcr{i}}$ are elements of an order $n$ tensor $\tsr{T}^{(\psi)} \in\mathbb{C}^{2 \times \cdots \times 2}$.
The \emph{inner product} of two quantum states $\ket{\psi}$ and $\ket{\phi}$ is given by
\[\braket{\phi | \psi} = \sum_{\vcr i \in \{0,1\}^n} ({t^{(\phi)}_{\vcr i}})^* t^{(\psi)}_{\vcr i},\]
and an \emph{operator} $\mat{H}$ acting on a quantum state is a linear transformation given by
\[\ket{\phi}=\mat{H}\ket{\psi}
 \quad \Rightarrow \quad t^{(\phi)}_{\vcr i} = \sum_{\vcr j \in \{0,1\}^n}  h_{\vcr i \vcr j} t^{(\psi)}_{\vcr{j}}.\]
The \emph{expectation value} of the Hermitian operator $\mat{H}$ for a quantum state $\psi$ is defined by \(\braket{\psi|\mat{H}|\psi}\), following the definitions of operators and inner products.

A quantum gate is an operator that can act on a small subset of qubits.
For example, a \emph{single qubit gate} $G^{(k)}$ acting on the $k$th qubit gives
\begin{align}\label{eq:one-site}
\ket{\phi} = G^{(k)} \ket{\psi} \:
\Rightarrow \: t^{(\phi)}_{\vcr i} = \sum_{j_k=0}^1 g^{(k)}_{i_kj_k} t^{(\psi)}_{i_1\cdots i_{k-1} j_k i_{k+1}\cdots i_n},
\end{align}
while a \emph{2-qubit gate} $G^{(k,l)}$ acting on qubits $k,l$ with $k<l$ gives
\begin{align}\label{eq:two-site}
\ket{\phi} = G^{(k,l)} \ket{\psi} \:
\Rightarrow \: t^{(\phi)}_{\vcr i} = \sum_{j_k=0}^1\sum_{j_l=0}^1 g^{(k,l)}_{i_ki_lj_kj_l} t^{(\psi)}_{i_1\cdots j_k \cdots j_l  \cdots i_n}.
\end{align}
These operator applications are particular examples of \emph{tensor contractions}, which allow for different types of products among tensors.

\subsection{Tensor Networks for Quantum States}

The contraction of a set of tensors defines a \emph{tensor network}, which could serve to provide a tensor decomposition~\cite{grasedyck2013literature,kolda2009tensor} of a quantum state.

The \emph{matrix product state} (MPS)~\cite{oseledets2011tensor,verstraete2008matrix} is a 1D tensor network that describes a quantum state $\psi$ with $n$ sites by $n$ tensors
$\tsr{M}^{(1)},\cdots\tsr{M}^{(n)}$, such that
\[
t^{(\psi)}_{i_1\cdots i_n}=
\sum_{k_1\cdots k_{n-1}}m^{(1)}_{i_1k_1}m^{(2)}_{i_2k_1k_2}\cdots m^{(n-1)}_{i_{n-1}k_{n-2}k_{n-1}}m^{(n)}_{i_nk_{n-1}}.
\]
By convention, $i_1,\cdots,i_n$ are called \emph{physical indices} and $k_1,\cdots,k_{n-1}$ are called \emph{bond indices}, while their dimensions are called \emph{physical dimensions} and \emph{bond dimensions} respectively.

An operator $\mat{H}$ may also be described as a 1D tensor network, which is known as the \emph{matrix product operator} (MPO)~\cite{khoromskij2011qtt,hubig2017generic,orus2014tnreview}:
\[
h_{i_1\cdots i_n j_1\cdots j_n}=
\sum_{k_1\cdots k_{n-1}}m^{(1)}_{i_1j_1k_1}m^{(2)}_{i_2j_2k_1k_2}\cdots
m^{(n)}_{i_nj_nk_{n-1}}.
\]

The \emph{projected entangled pair state} (PEPS)~\cite{verstraete2008matrix} is a 2D tensor network that represents a quantum state with $n^2$ sites on a square lattice by $n^2$ tensors in a way analogous to MPS. These $n^2$ tensors are arranged on
an $n\times n$ grid, so one can denote the tensor at the $p$'th row and $q$'th column
as $\tsr{M}^{(p,q)}$.
Then, each amplitude of a quantum state $\ket{\psi}$ is defined as
\[
t^{(\psi)}_{i_1\cdots i_{n^2}}=
\sum_{\vcr{k}}\prod_{pq}m^{(p,q)}_{i_{pn+q}\vcr{k}^{(p,q)}},
\]
where $\vcr{k}$ denotes all the indices shared by tensors in the PEPS and 
$\vcr{k}^{(p,q)}$ represents all the indices of the site tensor $\tsr{M}^{(p,q)}$ that
are shared with other site tensors. Similarly to MPS, indices $i_1\cdots i_{n^2}$ are called physical indices and $\vcr{k}^{(p,q)}$ are called bond indices.

\begin{figure}[htbp]
    \centering
    \includegraphics[width=0.9\columnwidth]{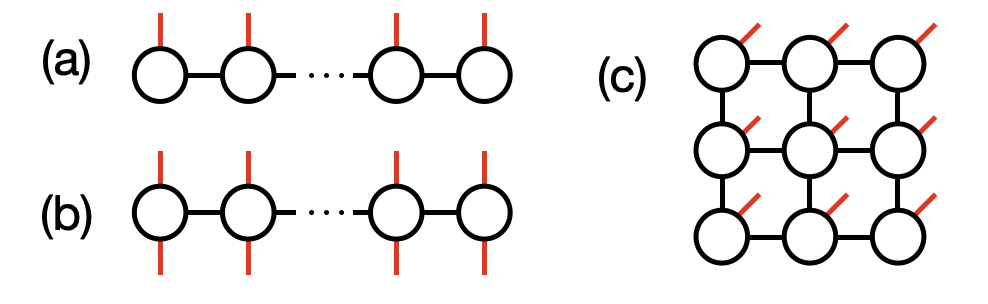}
    \caption{Tensor diagrams for (a) an MPS, (b) an MPO and (c) a $3 \times 3$ PEPS.}
    \label{fig:mps_mpo_peps}
\end{figure}

A tensor network can be visualized by a graph (\emph{tensor diagram}), in which each vertex denotes a tensor and each edge denotes an index.
Tensor diagrams describing MPS, MPO, and PEPS are displayed in Figure~\ref{fig:mps_mpo_peps}.

\subsection{Basic Tensor Network Computations}\label{subsec:tnc}

We now describe common computational primitives to transform tensor networks, focusing on PEPS.

\subsubsection{One-Site and Two-Site Operators}
To apply a one-site operator $\tsr{G}^{(1)}$ on the site $\tsr{M}^{(p,q)}$ of a PEPS, it suffices to compute the updated site $\tsr{\tilde{M}}^{(p,q)}$ by performing the contraction,
\begin{equation}\label{eq:peps-one-site}
\tilde{m}^{(p,q)}_{ik_1k_2k_3k_4} =\sum_{j}g^{(1)}_{ij}m^{(p,q)}_{jk_1k_2k_3k_4}.
\end{equation}
To apply a two-site operator $\tsr{G}^{(2)}$ on two neighboring sites $\tsr{M}^{(p,q)}$ and $\tsr{M}^{(p,q+1)}$, we consider a variant of the simple update algorithm~\cite{jiang2008su}. Two site tensors are contracted with the operator and then decomposed into two new site tensors $\tsr{\tilde{M}}^{(p,q)}$ and $\tsr{\tilde{M}}^{(p,q+1)}$,
\begin{align}\label{eq:peps-two-site}
&\:\sum_{k_4'}\tilde{m}^{(p,q)}_{i_1k_1k_2k_3k_4'}\tilde{m}^{(p,q+1)}_{i_2k_4'k_5k_6k_7}\nonumber\\
\approx\:&
\sum_{j_1j_2k_4} g^{(2)}_{i_1i_2j_1j_2}
m^{(p,q)}_{j_1k_1k_2k_3k_4}m^{(p,q+1)}_{j_2k_4k_5k_6k_7}.
\end{align}
This contraction and refactorization procedure is described by the first and last tensor diagrams in Figure~\ref{fig:qr-svd-apply}.
Applying a two-site operator on non-neighboring sites is possible by applying a chain of two-site operators (i.e. SWAP gates) on neighboring sites.

\begin{figure}[htbp]
\centering
\includegraphics[width=0.95\columnwidth]{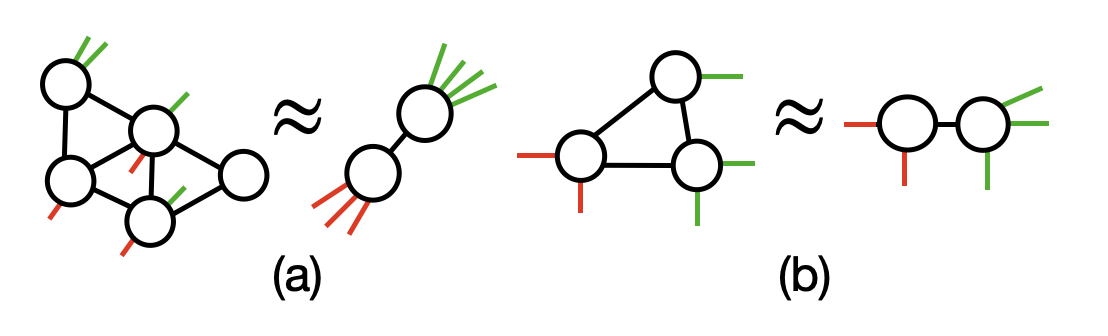}
\caption{\texttt{einsumsvd} examples: (a) a 5-site tensor network is contracted and then refactorized into two sites. (b) PEPS contraction step in which two tensors that share a neighbor are merged and the combined dimension is truncated.}
\label{fig:einsumsvd}
\end{figure}

The contraction and refactorization necessary to apply two-site operators may be executed by contracting the tensors (utilizing \texttt{einsum} in Python's NumPy), followed by reshaping the result into a matrix, computing a low-rank matrix factorization such as the truncated SVD, and reshaping the resulting factors into the appropriate tensors.
We refer to the combined refactorization operation as \texttt{einsumsvd}. This primitive takes as input a tensor network and produces a two-site tensor network with a single virtual leg connecting the two sites.
With the notion of \texttt{einsumsvd}, we can rewrite Equation~\eqref{eq:peps-two-site} as
\begin{align*}
\tsr{\tilde{M}}^{(p,q)},
\tsr{\tilde{M}}^{(p,q+1)}
\gets
\texttt{einsumsvd}(
\tsr{G},\tsr{M}^{(p,q)},\tsr{M}^{(p,q+1)}
).
\end{align*}
Figure~\ref{fig:einsumsvd} depicts other examples of \texttt{einsumsvd}.

\subsubsection{Amplitudes and Expectation Values}

A given amplitude of a state $\ket{\psi}$, such as $t^{(\psi)}_{\vcr i}$, may be obtained by computing $\braket{\vcr{i}|\psi}$. When represented by PEPS, the bond dimension of $\ket{\vcr i}$ is 1 and each component of $\ket{\vcr i}$ can be applied to the respective site of $\ket{\psi}$, resulting in a contraction of a PEPS without physical dimensions, i.e., a \emph{one-layer contraction}.

\begin{figure}[htbp]
\centering
\includegraphics[width=0.9\columnwidth]{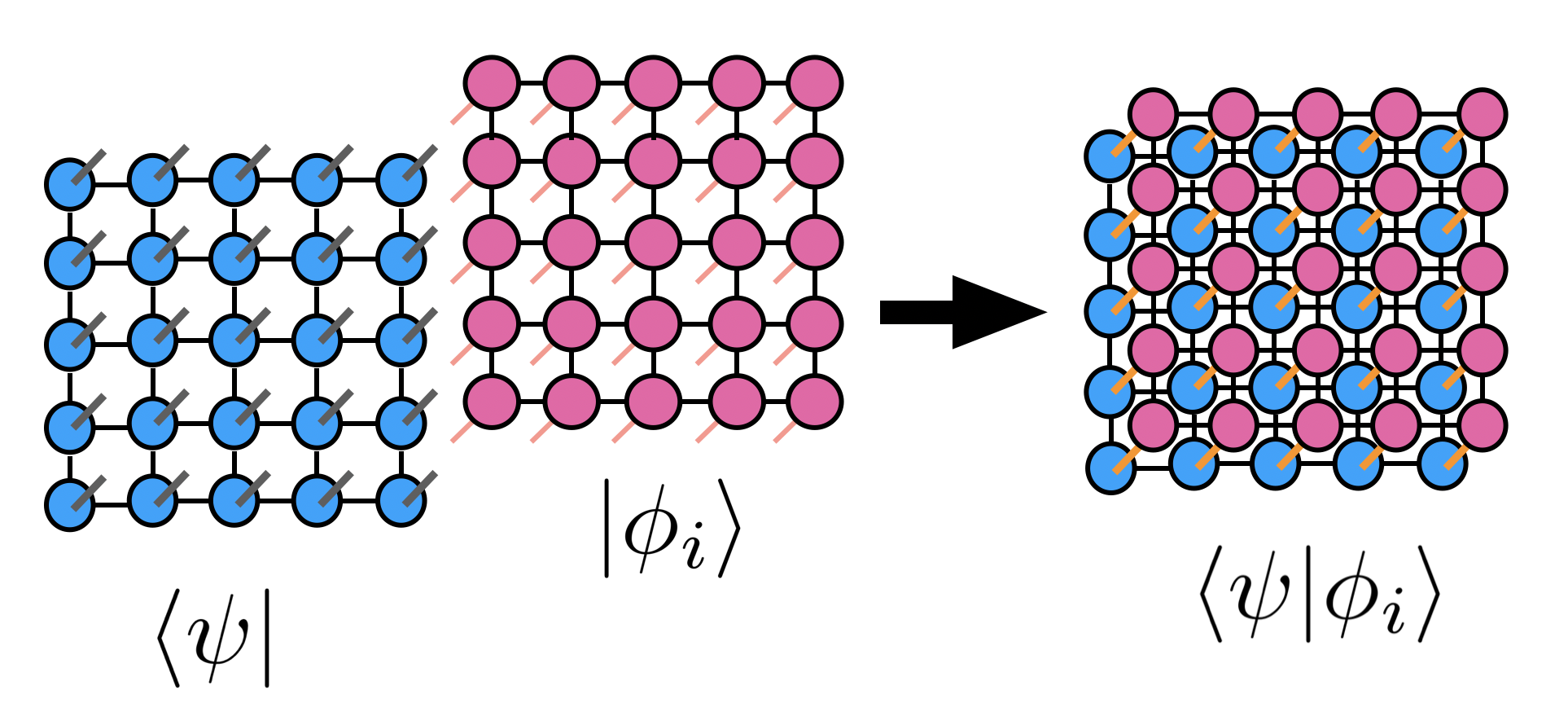}
\caption{Tensor diagram depicting the inner product of two PEPS networks.}
\label{fig:two-layer-contraction}
\end{figure}

Consider a Hermitian operator $\mat H$ composed of local terms $\mat{H}=\sum_i\mat{H}_i$, where $\mat{H}_i$ are operators acting on local sites of the PEPS. One way to calculate its expectation value is to calculate it as a sum of the expectation values of $\mat{H}_i$,
\begin{equation}\label{eq:peps-expectation}
\braket{\psi|\mat{H}|\psi}=\sum_{i=1}^{N}\braket{\psi|\mat{H}_i|\psi}=\sum_{i=1}^{N}\braket{\psi|\phi_i}.
\end{equation}
$\ket{\phi_i}$ can be obtained by the algorithms of applying local operators on $\ket{\psi}$. The calculation of $\braket{\psi|\phi_i}$ involves a contraction of two PEPS, i.e. a \emph{two-layer contraction}, as depicted in Figure~\ref{fig:two-layer-contraction}.

Both one-layer and two-layer contractions require exponential cost in the number of sites if done in an exact way.
A variety of approximate schemes with polynomial cost exists, which we survey in Section~\ref{section:prev-algo}.

\subsection{Tensor Network Applications}\label{subsec:tna}

Most tensor network computations are based on the operator application and contraction primitives introduced in the previous subsection. We provide a technical introduction to two representative methods below.

\subsubsection{Imaginary Time Evolution}\label{subsubsec:tebd}

Consider a Hamiltonian $\mat{H}=\sum_{j=1}^N \mat{H}_j$, where $\mat{H}_j$ are local operators. \emph{Imaginary time evolution} (ITE) evolves a quantum state toward its ground state by repetitively applying the ITE operator, $e^{-\tau\mat{H}}$, in effect computing power iteration on the matrix $e^{-\tau \mat{H}}$, which generally converges to the eigenvector with the smallest eigenvalue of $\mat H$.
ITE performs this iteration via the \emph{time-evolution block decimation} (TEBD) algorithm, which enables the approximate evolution of a tensor network state by decomposing $e^{-\tau\mat{H}}$ into a sequence of local operators using the Trotter-Suzuki decomposition:
\(e^{-\tau \mat{H}}=\prod_{j=1}^N e^{-\tau \mat{H}_j} + O(\tau^2).\)
Since each $\mat{H}_j$ is a one-site or two-site operator, so is $e^{-\tau \mat{H}_j}$.
Consequently, a Hamiltonian with $N$ local terms requires application of $N$ local operators for each ITE step.
The Rayleigh quotient $\frac{\braket{\psi|\mat{H}|\psi}}{\braket{\psi|\psi}}$ of the final state is computed as 
an approximation to the ground state energy.

\subsubsection{Quantum Circuit Simulation}
Following the prescription for gate application described in Section~\ref{subsec:qsat}, any quantum circuit can be expressed in terms of one-site and two-site operators acting on an initial state (e.g. a basis state).
Simulation of quantum circuits is useful for the development and verification of quantum algorithms and software. Approximate simulation additionally sheds insight on the propagation of error within quantum algorithms, which is useful for understanding the effect of noise on computations with near-term quantum devices~\cite{zhou2020limits}.
Among the quantum algorithms for near-term quantum devices, the \emph{variational quantum eigensolver} (VQE)~\cite{peruzzo2014vqe} is one of the most promising. 

VQE is a quantum-classical hybrid algorithm that computes the ground state of a Hamiltonian $\mat H$ by optimizing a quantum state parameterized by a quantum circuit.
The quantum part of the algorithm evaluates the objective $\braket{\psi(\vcr{\theta})|\mat{H}|\psi(\vcr{\theta})}$ for given parameters $\vcr{\theta}$, and the classical part of the algorithm takes the measurement results from the quantum device and runs classical optimization algorithms to tune the circuit parameters.
A careful design of the circuit structure, depending on the Hamiltonian of interest, is critical in improving the optimization efficiency and achieving desirable accuracy.
A number of efforts have been made to design more efficient circuits for different Hamiltonians~\cite{Liu_2019} and to apply the algorithm to quantum chemistry simulations~\cite{O'Malley2016scalable,grimsley2019adaptive,Parrish2019vqe}.

\section{Previous Work on PEPS Algorithms}
\label{section:prev-algo}

Our contributions build directly on existing algorithms for evolution and contraction of PEPS.

% The focus of our work is on accelerating tensor computations within the applications of PEPS.
% We now survey state-of-the-art techniques for the relevant primitives.
% These largely stem from physics literature, which is often not focused on comparative cost analysis and parallelization.

\subsection{PEPS Evolution Algorithms}\label{section:qrsvd}
\begin{figure}[htbp]
    \centering
    \includegraphics[width=0.95\columnwidth]{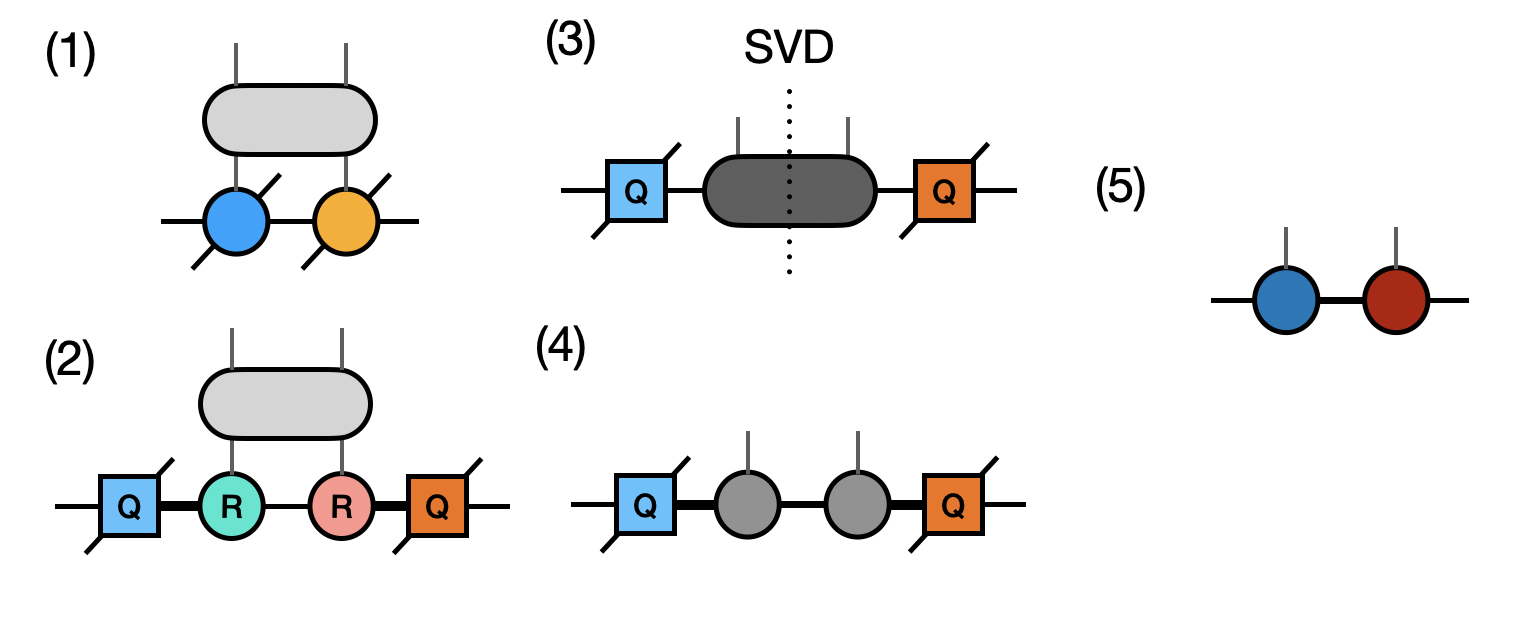}
    \caption{Tensor diagrams describing QR-SVD (Algorithm~\ref{algo:qr-svd-apply}).}
    \label{fig:qr-svd-apply}
\end{figure}

Direct evaluation of Equation~\ref{eq:peps-two-site} reveals a time complexity of $O(d^3r^9)$ and space complexity of $O(d^2r^6)$, where $d$ is the size of the physical dimension and $r$ is the size of the bond dimension.
Assuming $d\ll r$, the time and space complexity can be reduced to $O(d^2r^5)$ and $O(dr^4)$ respectively by a QR factorization of the two site tensors prior to the application of the operator\cite{zhou2020limits}, as shown in Algorithm~\ref{algo:qr-svd-apply} and Figure~\ref{fig:qr-svd-apply}.

\begin{algorithm}[htbp]
\caption{Operator application by QR-SVD}\label{algo:qr-svd-apply}
\begin{algorithmic}[1]
\REQUIRE two site tensors $\tsr{M}^{(p,q)}$ and $\tsr{M}^{(p,q+1)}$,\\
    a two-site operator $\tsr{G}$
\ENSURE updated site tensors $\tsr{\tilde{M}}^{(p,q)}$ and $\tsr{\tilde{M}}^{(p,q+1)}$
\STATE $\tsr{Q}^{(p,q)},\tsr{R}^{(p,q)}\gets\text{QR}(\tsr{M}^{(p,q)})$
\COMMENT{step (1) $\rightarrow$ (2)}
\STATE $\tsr{Q}^{(p,q+1)},\tsr{R}^{(p,q+1)}\gets\text{QR}(\tsr{M}^{(p,q+1)})$
\COMMENT{step (1) $\rightarrow$ (2)}
\STATE $\tsr{\tilde{R}}^{(p,q)},\tsr{\tilde{R}}^{(p,q+1)}
\gets\texttt{einsumsvd}(\tsr{G},\tsr{R}^{(p,q)},\tsr{R}^{(p,q+1)})$\\
\COMMENT{step (2) $\rightarrow$ (4)}
\STATE $\tsr{\tilde{M}}^{(p,q)}\gets\tsr{Q}^{(p,q)}\tsr{\tilde{R}}^{(p,q)}$
\COMMENT{step (4) $\rightarrow$ (5)}
\STATE $\tsr{\tilde{M}}^{(p,q+1)}\gets\tsr{Q}^{(p,q+1)}\tsr{\tilde{R}}^{(p,q+1)}$
\COMMENT{step (4) $\rightarrow$ (5)}
\RETURN $\tsr{\tilde{M}}^{(p,q)}$, $\tsr{\tilde{M}}^{(p,q+1)}$
\end{algorithmic}
\end{algorithm}

\subsection{PEPS Contraction Algorithms}

% To extract information from quantum states in the PEPS representation, a key primitive is the contraction of a PEPS into a scalar.
% %fundamental step is to contract the tensor network formed by two PEPS with corresponding free index pairs at each site connected.
% Contracting a tensor network formed by two PEPS with corresponding free index pairs at each site-connected tensor network gives the inner product between two PEPS-represented quantum states $\braket{\phi|\psi}$, as illustrated in Figure \ref{fig:two-layer-contraction}.
% %For expectation values, where $\phi$ is the resulting state of applying multiple single site and local two site gates to the initial state and $\psi$ is the eigenstate guess we want to compute the Rayleigh-quotient of, as illustrated in Figure \ref{fig:contract_background}.
% As exact contraction involves handling intermediate tensors of exponentially large sizes, it becomes the bottleneck of simulating quantum systems classically. The most memory-efficient exact contraction algorithm \cite{guo2019simulation}, while efficient for very small bond dimension, has exponential cost in the number of sites of a square PEPS. As a result,

Many approximate PEPS contraction algorithms have been proposed in pursuit of reducing memory footprint while maintaining high accuracy. These include the original Boundary MPS (BMPS)\cite{verstraete2004peps} algorithm and its variants \cite{pivzorn2011time, lubasch2014unifying, lubasch2014algorithms}, Tensor Renormalization Group (TRG) \cite{levin2007trg} and many TRG-inspired renormalization group methods \cite{gu2008terg, gu2009tefr, xie2009srg, zhao2010srg, xie2012hotrg, evenbly2015tnr, zhao2016hosrg, shuo2017loop-tnr}, Corner Transfer Matrix (CTM) \cite{nishino1996ctm, orus2009ctm}, as well as contraction algorithms for new 2D tensor network structures based on PEPS \cite{xie2014pess, xie2017ntn}.
We focus on a variant of BMPS for one-layer and two-layer contractions as introduced below.
% We focus on BMPS,one of the most important algorithms that can be applied to 2D finite PEPS with a square lattice. We consider versions of BMPS for one-layer and two-layer contraction.

\subsubsection{One-Layer BMPS}
    % Given two $n$ by $n$ PEPS $\tsr{A}$ and $\tsr{B}$ of bond dimension $r_1$ and $r_2$ respectively,
    % BMPS algorithm can be conducted in two ways. The naive way is to contract over all the physical indices between $\tsr{A}$ and $\tsr{B}$ first to create a new $n$-by-$n$ PEPS $\tsr{P}$ of bond dimensions $r=r_1r_2$ without physical indices, and perform One-layer BMPS on $\tsr{P}$.
    Let $\tn{P}$ be an $n$-by-$n$ PEPS of bond dimension $r$ without physical indices and $\tn{P}^{(1)},\cdots,\tn{P}^{(n)}$ be the rows of $\tn{P}$. Treating $\tn{P}^{(1)}$ as an MPS and $\tn{P}^{(2)},\cdots, \tn{P}^{(n)}$ as MPOs, contracting $\tn P$ by BMPS can be considered as applying MPOs $\tn{P}^{(2)},\cdots, \tn{P}^{(n)}$ on the MPS $\tn{P}^{(1)}$ approximately. The approximation is necessary
    for avoiding exponential cost and it is done by truncating the bond dimension of result MPS to a predefined $m$. This approach is described in Algorithm~\ref{algo:bmps}.

    \begin{algorithm}[htbp]
        \caption{Boundary MPS}\label{algo:bmps}
        \begin{algorithmic}[1]
        \REQUIRE $n\times n$ PEPS $\tn{P}$ consisting of rows $\{\tn{P}^{(1)},\cdots,\tn{P}^{(n)}\}$,\\
            algorithm \texttt{ApproxApply},
            truncation bond dimension $m$
        \ENSURE scalar $v$
        \STATE $\tn{S}\gets$ $\tn{P}^{(1)}$
        \FOR {$i=2,3,\dots,n$}
        \STATE $\tn{S}\gets\texttt{ApproxApply}(\tn{S},\tn{P}^{(i)}, m)$
        \ENDFOR \\
        \COMMENT{$\tn{S}$ is an MPS without physical indices at this point}
        \STATE $v\gets$ contracting $\tn{S}$ to a scalar
        \RETURN $v$
        \end{algorithmic}
    \end{algorithm}
    
\begin{figure}[htbp]
        \centering
        \includegraphics[width=0.9\columnwidth]{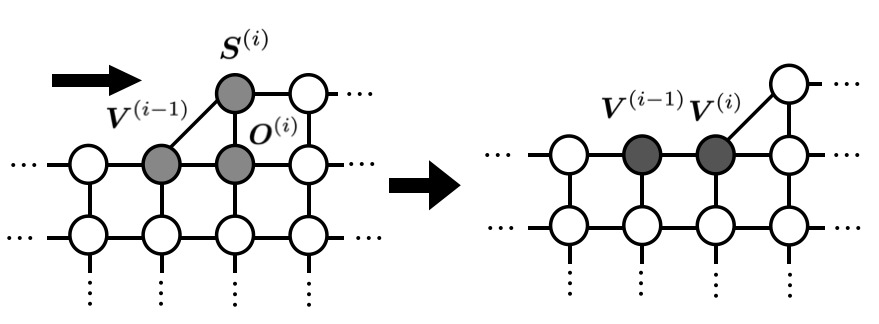}
        \caption{Apply MPO on MPS approximately (Algorithm~\ref{algo:approx-apply}).}
        \label{fig:bmps}
    \end{figure}

    The approximate application of an MPO on an MPS can be performed in various ways. The original BMPS algorithm \cite{verstraete2004peps} involves solving a linear equation $\tn{E}\tsr{T}=\tn{P}$ for one site $\mat{T}$ while fixing the other sites (the environment $\tn{E}$) at a time with respect to the target MPS $\tn{P}$. This method, while having a low cost of $O(dm^2r^3)+O(m^3r^2)$ if the MPO has a two-layer structure as discussed below \cite{lubasch2014algorithms},
    is typically done with multiple rounds of alternating optimization of the sites of the MPS.
    Our approach is most closely related to
    the zip-up algorithm proposed in \cite{stoudenmire2010compress}, where a series of \texttt{einsumsvd}s are conducted to truncate the bond dimension of $\tn{P}$, as shown in Figure~\ref{fig:bmps} and described in Algorithm \ref{algo:approx-apply}.
    We also note that this method can incorporate canonicalization (orthogonalization of the tensor network) to minimize amplification of SVD truncation error.
    Specifically, canonicalization can be achieved by absorbing singular values into $\tsr{V}^{(i-1)}$ or $\tsr{V}^{(i)}$ for different rows of PEPS \cite{stoudenmire2010compress} in an alternating manner.

    % \sout{This variant does not require sweeps, but
    % if implemented naively (i.e. contracting three tensors and performing SVD in the \texttt{einsumsvd} step), would have a higher cost of $O(m^3r^4)$.
    % In Section~\ref{section:algorithms}, we discuss one way to reduce this cost to $O(m^3r^2)$.}

    \begin{algorithm}[htbp]
        \caption{Apply MPO on MPS approximately}\label{algo:approx-apply}
        \begin{algorithmic}[1]
        \REQUIRE MPS $\tn{S}$ consisting of tensors $\{\tsr{S}^{(1)},\cdots, \tsr{S}^{(n)}\}$,\\
            MPO $\tn{O}$ consisting of tensors $\{\tsr{O}^{(1)},\cdots, \tsr{O}^{(n)}\}$,\\
            algorithm \texttt{einsumsvd}, truncation bond dimension $m$
        \ENSURE MPS $\tn{S'}$
        \STATE $\tsr{V}^{(1)}\gets$ Contract $\tsr{S}^{(1)}$ and $\tsr{O}^{(1)}$
        \COMMENT {$v^{(1)}_{jkl}\gets \sum_{i} s^{(1)}_{ij}o^{(1)}_{ikl}$}
        \FOR {$i=2,3,\dots,n$}
        \STATE $\tsr{V}^{(i-1)},\tsr{V}^{(i)}\gets\texttt{einsumsvd}(\tsr{V}^{(i-1)},\tsr{S}^{(i)},\tsr{O}^{(i)},m)$
        \ENDFOR
        \STATE $\tn{S'}\gets \{\tsr{V}^{(1)},\cdots,\tsr{V}^{(n)}\}$
        \RETURN $\tn{S'}$
        \end{algorithmic}
    \end{algorithm}

\subsubsection{Two-Layer BMPS}
    % Two-layer contraction can be reduced to a one-layer contraction by combining respective sites from the two PEPS layers.
    Given two $n$-by-$n$ PEPS $\tn{A}$ and $\tn{B}$ of bond dimension $r_1$ and $r_2$ respectively,
    their inner product can be computed in two ways. The naive way is to contract over all the physical indices between $\tn{A}$ and $\tn{B}$ to create a new $n$-by-$n$ PEPS $\tn{P}$ with bond dimensions $r=r_1r_2$ and no physical dimensions, and then perform one-layer BMPS.
    However, forming the combined $\tn{P}$ requires $O(r_1^4r_2^4)$ memory, while storing $\tn{A}$ and $\tn{B}$ separately only needs $O(r_1^4+r_2^4)$.
    An improvement is to maintain the two-layer structure and only contract the corresponding sites from two layers to one layer when needed, which is numerically equivalent to the naive way but saves memory. This approach is often used in practice and referred to interchangeably with the naive approach in the literature. In Section~\ref{section:algorithms}, we show that when \texttt{einsumsvd} is equipped with implicit randomized SVD, the implicit structure of the two-layer BMPS approach yields reduced computational complexity.

% \section{New Algorithms for PEPS Simulation}
\section{Algorithmic Improvements to PEPS Simulation}\label{section:algorithms}

In this section, we describe our algorithmic improvements to basic PEPS computations, specifically \texttt{einsumsvd} and the expectation value calculation introduced in Section~\ref{subsec:tnc}.

% We propose new methods for accelerating the tensor computations arising in PEPS applications.
% We first describe a new method to exploit reuse among multiple different PEPS contractions.
% Then, we propose a more efficient method for contracting and refactorizing subnetworks during PEPS contraction.
% Both of these optimizations alleviate the cost of the most expensive component of PEPS simulation for methods like ITE and VQE.

\subsection{Implicit Tensor Network Refactorization}\label{subsec:ibmps}

\begin{table*}[htbp]
    \centering
    \begin{tabular}{| m{8em} | m{10em} | m{10em} | m{12em} |}
        \hline
    	& BMPS & IBMPS & Two-layer IBMPS \\
    	\hline 
    	\text{Time complexity} & $O(n^2m^3r^4)$ & $O(n^2m^2r^4+n^2m^3r^2)$ & $O(n^2dm^2r^3+n^2dm^3r^2)$\\
    	\hline 
    	\text{Space complexity} & $O(\max\{m^2r^3,r^4\})$ & $O(\max\{m^2r^2,r^4\})$ & $O(\max\{m^2r^2,r^4\})$\\
    	\hline 
    \end{tabular}
    \caption{The asymptotic time and space complexity of computing $\braket{\tn{P}|\tn{P}}$
    using BMPS, IBMPS and two-layer IBMPS
    for an $n$-by-$n$ PEPS $\tn{P}$ of bond dimension $\sqrt{r}$, physical dimension $d$, and truncation bond dimension $m$.}
    % the table compares (introduced in Section~\ref{subsec:ibmps}).
    \label{tab:ctr_costs}
\end{table*}

Randomized SVD is an algorithm that approximates truncated SVD of rank $r$ of
an $m$-by-$n$ matrix with time complexity $O(mnr)$ \cite{halko2011randomness} by the use of orthogonal iteration.
This algorithm has been directly applied to replace the full SVD algorithm to apply operators on MPS\cite{tamascelli2015rsvd}, which reduces the cost by one order of the physical dimension. Here we apply this idea more generally to \texttt{einsumsvd}, resulting in asymptotic reductions in cost for one-layer and two-layer BMPS contraction.

\begin{algorithm}[htbp]
\caption{Randomized SVD}\label{algo:randomized-svd}
\begin{algorithmic}[1]
\REQUIRE
    an operator $\tsr{A}:\C^{n_1\times\cdots\times n_t}\mapsto\C^{m_1\times\cdots\times m_s}$,
    rank $r$,
    number of iteration $k$
\ENSURE an approximated truncated SVD of the input operator\\
    tensor operator $\tsr{U}\in\C^{m_1\times\cdots\times m_s\times r}$: $\C^{r}\mapsto\C^{m_1\times\cdots\times m_s}$,\\
    diagonal matrix $\mat{\Sigma}\in\C^{r\times r}$,\\
    tensor operator $\tsr{V}\in\C^{r\times n_1\times\cdots\times n_t}$: $\C^{r}\mapsto\C^{n_1\times\cdots\times n_t}$
\STATE $\tsr{Q}\gets$ draw a random tensor from $[-1,1]^{n_1\times\cdots\times n_t\times r}$\\
\COMMENT {$\tsr{Q}: \C^{r}\mapsto\C^{n_1\times\cdots\times n_t}$}
\STATE $\tsr{P}\gets\text{orthogonalize}(\tsr{A}\tsr{Q})$
\COMMENT {$\tsr{P}: \C^{r}\mapsto\C^{m_1\times\cdots\times m_s}$}
\FOR {$i=1,2,\dots,k$}
\STATE $\tsr{Q}\gets\text{orthogonalize}(\tsr{A}^*\tsr{P})$
\COMMENT {$\tsr{Q}: \C^{r}\mapsto\C^{n_1\times\cdots\times n_t}$}
\STATE $\tsr{P}\gets\text{orthogonalize}(\tsr{A}\tsr{Q})$
\COMMENT {$\tsr{P}: \C^{r}\mapsto\C^{m_1\times\cdots\times m_s}$}
\ENDFOR
\STATE $\tilde{\tsr{U}},\mat{\Sigma},\tsr{V}\gets\text{SVD}(\tsr{P}^*\tsr{A})$
\COMMENT {$\tsr{P}^*\tsr{A}$: $\C^{n_1\times\cdots\times n_t}\mapsto\C^{r}$}
\STATE $\tsr{U}\gets\tsr{P}\tilde{\tsr{U}}$
\RETURN $\tsr{U},\mat{\Sigma},\tsr{V}$
\end{algorithmic}
\end{algorithm}

As shown in Algorithm~\ref{algo:randomized-svd}, randomized SVD does not require
an explicit form of the operator $\tsr{A}$ but only needs to know how to apply
$\tsr{A}$ and $\tsr{A}^*$. Thus, the operator $\tsr{A}$ could be given implicitly
as an uncontracted tensor network. Depending on the structure of the tensor network
$\tsr{A}$, it is possible to perform $\tsr{A}\tsr{Q}$ and $\tsr{A}^*\tsr{P}$ more
efficiently and with less memory than by explicitly forming $\tsr{A}$ and applying it.

A similar implicit method has been specifically applied to the TRG algorithm~\cite{morita2018tensor}.
Here, we apply this method to the \texttt{einsumsvd} step of the BMPS contraction algorithm (Algorithm~\ref{algo:approx-apply}).
With the best choice of contraction order, the use of implicit methods in BMPS yields an asymptotic reduction in cost and, in some cases, memory footprint, when compared with a naive implementation of \texttt{einsumsvd}.
We refer to this approach as \emph{implicit} BMPS (IBMPS) contraction.
Two-layer BMPS can achieve greater benefit with the use of implicit randomized SVD, as it naturally maintains tensors in implicit forms.
We compare the asymptotic costs for BMPS, IBMPS, and two-layer IBMPS in Table~\ref{tab:ctr_costs}.

\subsection{Intermediate Caching for PEPS Expectation Values}\label{subsec:intermediate-caching}

To compute expectation values of PEPS efficiently, we develop a caching strategy for consecutive executions of BMPS (Algorithm~\ref{algo:bmps}) when evaluating Equation~\eqref{eq:peps-expectation}. We observe that these contractions share common intermediates due to the locality of $\mat{H}_i$, and our approach reuses partial contractions of rows (or columns) of PEPS to compute $\braket{\psi|\mat{H}_i|\psi}$.

\begin{figure}[htbp]
    \centering
    \includegraphics[width=0.7\columnwidth]{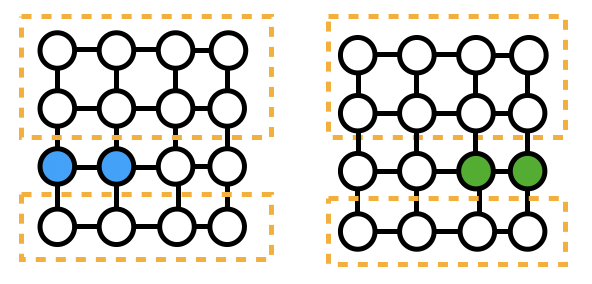}
    \caption{Shared intermediates for expectation values of two local operators on a $4\times 4$ PEPS.}
    \label{fig:expectation-cache}
\end{figure}

For example, consider a $4\times 4$ PEPS and two local operators $\mat{H}_1$ and $\mat{H}_2$ both acting on the third row of the PEPS, as shown in Figure~\ref{fig:expectation-cache}. We cache the two boundary MPS that represent the partial contraction results of the first two rows
and last row to accelerate the calculation of $\braket{\psi|\mat{H}_1|\psi}$ and $\braket{\psi|\mat{H}_2|\psi}$.

When applied to an $n\times n$ PEPS, this caching strategy requires two full two-layer PEPS contractions to generate all the cached intermediates, and then each local-operator expectation value is calculated with a $3\times n$ PEPS contraction. This strategy can be easily extended to compute a collection of expectation values, e.g., to compute gradients of parameters of local operators (as used in the Adapt-VQE method~\cite{grimsley2019adaptive}).
In Section~\ref{section:results}, we demonstrate the speed-up achieved by this strategy (see Figure~\ref{fig:expectation-cache-time}).

% The caching strategy lowers the cost of the expectation value calculation of $\mat{H}$ by a factor of $\Theta(N)$ and is easily extended to compute a collection of expectation values, e.g., to compute gradients of parameters of local operators (as used in the Adapt-VQE method~\cite{grimsley2019adaptive}).

Besides evaluating Equation~\eqref{eq:peps-expectation} with intermediate caching, $\braket{\psi|\mat{H}|\psi}$ can be calculated in an alternative approach. 
%For example, we can use
Using the Trotter-Suzuki decomposition,
\(e^{\tau \mat{H}}=\prod_{j=1}^N e^{\tau \mat{H}_j} + O(\tau^2)\),
as in ITE, and the Taylor expansion,
\(e^{\tau\mat{H}}=\mat{I}+\tau\mat{H}+ O(\tau^2)\),
one can approximate
\begin{equation}\label{eq:trotter-expval}
\braket{\psi|\mat{H}|\psi}=\frac{1}{\tau}\left(\braket{\psi|\prod_{j=1}^N e^{\tau\mat{H}_j}|\psi}-\braket{\psi|\psi}\right) + O(\tau).
\end{equation}
%The intermediate term $\ket{\phi}=\prod_{j=1}^N e^{t\mat{H}_j}\ket{\psi}$ in Equation~\eqref{eq:trotter-expval} can be evaluated by applying $N$ local operators on $\ket{\psi}$ in a similar way as of a single ITE step (as introduced in next subsection).
In comparison with the previous method,
Equation~\eqref{eq:trotter-expval} requires one two-layer PEPS contraction as opposed to two, but application of an additional ITE step would grow the bond dimension of $\ket{\phi}=\prod_{j=1}^N e^{t\mat{H}_j}\ket{\psi}$ or require approximation.
%is generally a few times larger than the bond dimension of $\ket{\phi_i}$ in Equation~\eqref{eq:peps-expectation}. Hence, 
%The overall costs are comparable due to the complexity of PEPS contraction (Table~\ref{tab:ctr_costs}).

% By exploiting reuse within each row or column in the PEPS, we amortize the cost of contracting the boundary.
% Previous work on BMPS exploited reuse of terms for particular Hamiltonians~\cite{verstraete2004peps}, leveraging alternating optimization within each PEPS row. Our scheme is easier to parallelize effectively, as optimization requires multiple sweeps over the sites.

\section{Koala: High-Performance PEPS Simulation} \label{section:implementation}
We implement the algorithms that are discussed above within an open-source Python library
called ``Koala'', which is available at \url{https://github.com/cyclops-community/koala}.

\subsection{Interface and Features}

Koala provides explicit primitives for constructing PEPS networks, applying operators, and computing expectation values.
Tensor library backends, as well as different algorithms for applying operators and contracting PEPS, can be specified as arguments to these primitive routines. An example code for constructing a PEPS, applying operators, and computing an expectation value is below.
\begin{lstlisting}[language=Python,captionpos=b,label={lst:koala}]
from koala import peps, Observable
from koala.peps import QRUpdate, BMPS
from tensorbackends.interface import \
  ImplicitRandomizedSVD

# Create a 2-by-3 PEPS in distributed memory
qstate = peps.computational_zeros(
  nrow=2, ncol=3, backend='ctf'
)

# Apply one-site and two-site operators with QR-SVD
Y = qstate.backend.astensor([...]).reshape(2,2)
CX = qstate.backend.astensor([...]).reshape(2,2,2,2)
qstate.apply_operator(Y, [1]) 
qstate.apply_operator(CX, [1,4], QRUpdate(rank=2))

# Calculate the expectation value with IBMPS
H = Observable.ZZ(3, 4) + 0.2 * Observable.X(1)
result = qstate.expectation(
  H, use_cache=True,
  contract_option=BMPS(ImplicitRandomizedSVD(rank=4))
)
\end{lstlisting}

\vspace*{-.3in}

\subsection{Distributed Tensor Parallelization}

Koala supports both sequential and parallel execution. This portability is achieved by an abstraction layer of tensor data types and operations that enables different tensor libraries to be used with little change in the source code. Currently, the tensor libraries we support
are NumPy, CuPy, and Cyclops. Among them, NumPy provides sequential and threaded routines for tensor computations~\cite{oliphant2006numpy}, CuPy implements such routines that run on GPUs~\cite{nishino2017cupy}, and Cyclops allows similar operations in distributed memory~\cite{solomonik2014massively}. Cyclops distributes each tensor over all processors and provides a Python interface for \texttt{einsum} operations, which can specify the contraction of any tensor network into a single tensor. Cyclops also provides a front-end to routines in the ScaLAPACK library~\cite{blackford1997scalapack} for distributed matrix factorization.

\subsection{Avoiding Tensor Reshaping}

The standard approach to \texttt{einsumsvd} involves performing tensor contractions, matricizing (unfolding) the result, computing a low-rank matrix factorization, and folding the factors into tensors of appropriate shape.
Folding and unfolding of this type are logical transformations that need not modify the data and often have negligible cost in sequential execution.
However, these two operations can be nontrivial if each tensor is distributed on a processor grid (via any standard distribution, including cyclic and blocked layouts).
Since both the matrix and the tensor unfoldings must be mapped to different processor grids for corresponding computations, folding/unfolding requires expensive redistribution of data. Therefore, while calls to NumPy's \texttt{reshape} are practically free, Cyclops' \texttt{reshape} can become a bottleneck.

We address this bottleneck within implicit \texttt{einsumsvd} (Algorithm~\ref{algo:randomized-svd}) and QR-SVD (Algorithm~\ref{algo:qr-svd-apply}), by performing orthogonalization via eigendecomposition of a Gram matrix formed by tensor contraction.
For the QR factorizations done in QR-SVD, forming the Gram matrix requires a reshape of a small tensor.
For orthogonalization in \texttt{einsumsvd}, no reshape of the Gram matrix is necessary.
In both cases, this matrix is small, so the latency overhead of using distributed memory is alleviated by performing computations on the Gram matrix sequentially. Algorithm~\ref{algo:reshap-avoiding-orthogonalization} describes this general approach.
When Algorithm~\ref{algo:reshap-avoiding-orthogonalization} is applied to QR-SVD, we can also perform \texttt{einsumsvd} for operator application sequentially by leveraging the fact that the operator $\tsr{G}$ and the $\mat{R}$ factors are all small enough to fit into local memory.

{%\small
\begin{algorithm}[bt]
\caption{Reshape-avoiding orthogonalization}\label{algo:reshap-avoiding-orthogonalization}
\begin{algorithmic}[1]
\REQUIRE
    a distributed tensor $\tsr{A}\in\C^{m_1\times\cdots\times m_s\times n_1\times\cdots\times n_t}$
    such that $\prod_{i=1}^{s}m_i \gg \prod_{i=1}^{t}n_i$\\
    ($\tsr{A}$ represents an operator
    $\C^{n_1\times\cdots\times n_t}\mapsto\C^{m_1\times\cdots\times m_s}$)
\ENSURE \quad\\
    distributed isometric tensor operator $\tsr{Q}\in\C^{m_1\times\cdots\times m_s\times n_1\times\cdots\times n_t}$:\\
    \quad $\C^{n_1\times\cdots\times n_t}\mapsto\C^{m_1\times\cdots\times m_s}$\\
    distributed tensor operator $\tsr{R}\in\C^{n_1\times\cdots\times n_t\times n_1\times\cdots\times n_t}$:\\
    \quad $\C^{n_1\times\cdots\times n_t}\mapsto\C^{n_1\times\cdots\times n_t}$\\
\STATE In parallel, Compute $\tsr{G}\gets \tsr{A}^*\tsr{A}$ \\
\COMMENT {$\tsr{G}: \C^{n_1\times\cdots\times n_t}\mapsto\C^{n_1\times\cdots\times n_t}$}
\STATE Gather $\tsr{G}$ data among processors
\STATE Reshape $\tsr{G}$ into a matrix $\mat{G}\in\C^{n_1\cdots n_t \times n_1\cdots n_t}$
\STATE Compute eigendecomposition $\mat{X}$, $\mat{\Lambda}$ of $\mat{G}$ such that\\
    \quad $\mat{G}=\mat{X}\mat{\Lambda}\mat{X}^*$
\STATE Compute $\mat{R}\gets\sqrt{\mat{\Lambda}}\mat{X}^*\in\C^{n_1\cdots n_t \times n_1\cdots n_t}$
\STATE Compute $\mat{P}\gets\mat{R}^{-1}\in\C^{n_1\cdots n_t \times n_1\cdots n_t}$
\STATE Reshape $\mat{R}$ into a tensor $\tsr{R}\in\C^{n_1\times\cdots\times n_t\times n_1\times\cdots\times n_t}$\\
\COMMENT {$\tsr{R}: \C^{n_1\times\cdots\times n_t}\mapsto\C^{n_1\times\cdots\times n_t}$}
\STATE Reshape $\mat{P}$ into a tensor $\tsr{P}\in\C^{n_1\times\cdots\times n_t\times n_1\times\cdots\times n_t}$\\
\COMMENT {$\tsr{P}: \C^{n_1\times\cdots\times n_t}\mapsto\C^{n_1\times\cdots\times n_t}$}
%\STATE Send $\tsr{R}$ and $\tsr{P}$ to distributed memory
%\STATE $\tsr{Q}\gets\tsr{A}\tsr{P}$ in distributed memory
\STATE Distribute $\tsr{R}$ and $\tsr{P}$ among different processors
\STATE In parallel, compute $\tsr{Q}\gets\tsr{A}\tsr{P}$
\RETURN $\tsr{Q},\tsr{R}$
\end{algorithmic}
\end{algorithm}
}

\section{Results} \label{section:results}

We benchmark our implementation of the PEPS algorithm variants that are discussed above. The overall accuracy of PEPS-based simulation for two example applications is also studied to demonstrate how accuracy is affected by the bond dimension.
Parallel experiments are all done on the Stampede2 supercomputer using 64 cores per node for Cyclops and 32 MKL~\cite{wang2014intel} threads on one node for NumPy.
The number of \emph{processes per node} (PPN) for Cyclops is always chosen as 64 unless otherwise specified.

\subsection{PEPS Evolution Benchmark}
\begin{figure}[htbp]
\centering
     \begin{subfigure}[ht]{0.95\columnwidth}
         \includegraphics[width=0.95\columnwidth]{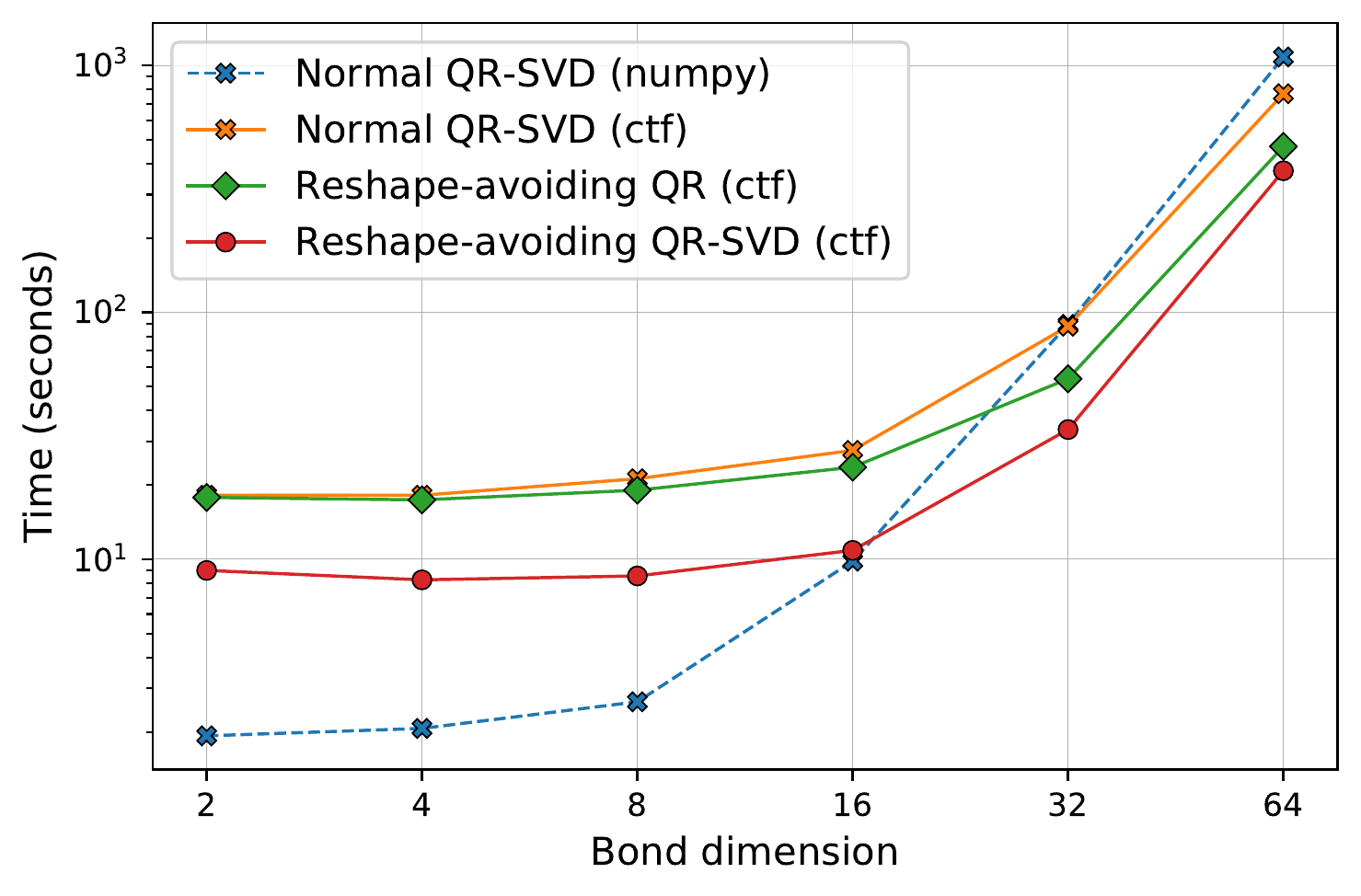}
         \caption{$8 \times 8$ PEPS using $1$ node.}
         \label{fig:compare_updates_8}
     \end{subfigure}
     \begin{subfigure}[ht]{0.95\columnwidth}
         \includegraphics[width=0.95\columnwidth]{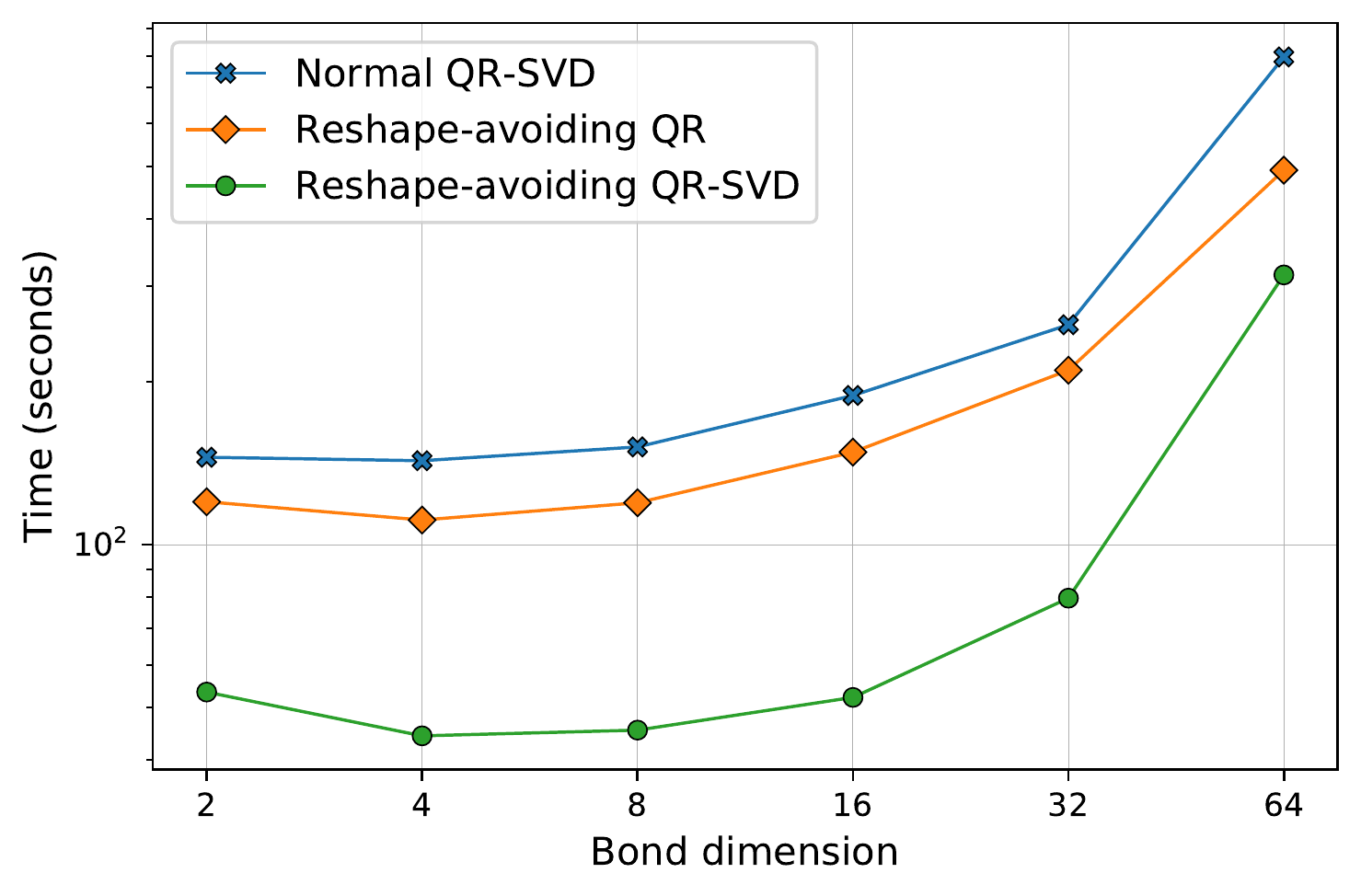}
         \caption{$15 \times 15$ PEPS using $16$ nodes.}
         \label{fig:compare_updates_15}
     \end{subfigure}
    \caption{Running time versus bond dimension for applying one-layer TEBD on PEPS using variants of QR-SVD. ``Reshape-avoiding QR'' denotes the variant where orthogonalization is done locally; ``reshape-avoiding QR-SVD'' denotes the variant where both orthogonalization and \texttt{einsumsvd} are done locally.}\label{fig:compare_updates_a}
    \label{fig:compare_updates}
\end{figure}

Figure~\ref{fig:compare_updates} studies the performance of applying one layer of TEBD operators on all neighboring sites on PEPS of two different sizes: $8 \times 8$ and $15 \times 15$. For the smaller system size, which fits into memory on $1$ node, we compare the performance of NumPy and Cyclops, as shown in Figure~\ref{fig:compare_updates_8}.
We observe that while NumPy outperforms Cyclops for small bond dimensions, their performance becomes similar as bond dimension grows. We also compare three different algorithms for a $15 \times 15$ PEPS on $16$ nodes using Cyclops in Figure~\ref{fig:compare_updates_15}, which shows
that reshape-avoiding orthogonalization (Algorithm~\ref{algo:reshap-avoiding-orthogonalization}) can accelerate the parallel execution of QR-SVD (Algorithm~\ref{algo:qr-svd-apply}) by factors of up to 3.6X.
%\sout{Figure~\ref{fig:compare_updates} studies the performance of
%applying one layer TEBD operators with NumPy and Cyclops backend on one node.
%We show that while NumPy outperforms Cyclops for small bond dimensions,
%their performance gets close as bond dimension grows.
%Moreover, this figure demonstrates that reshape-avoiding orthogonalization (Algorithm~\ref{algo:reshap-avoiding-orthogonalization}) improves the performance of Cyclops.}

% \begin{figure*}[htbp]
%     \centering
%     \begin{subfigure}{0.48\textwidth}
%         \includegraphics[width=\textwidth]{figures/compare_updates_8x8.pdf}
%         \caption{$8 \times 8$ PEPS with $1$ node}\label{fig:compare_updates_a}
%     \end{subfigure}
%         \begin{subfigure}{0.48\textwidth}
%         \includegraphics[width=\textwidth]{figures/compare_updates_15x15.pdf}
%     \caption{$15 \times 15$ PEPS with $16$ nodes}\label{fig:compare_updates_b}
%     \end{subfigure}
%     \caption{Running time for one layer TEBD versus the rank.}\label{fig:compare_updates}
% \end{figure*}

\subsection{PEPS Contraction Benchmark}

\begin{figure}[htbp]
\centering
\begin{subfigure}[t]{0.95\columnwidth}
    \includegraphics[width=\columnwidth]{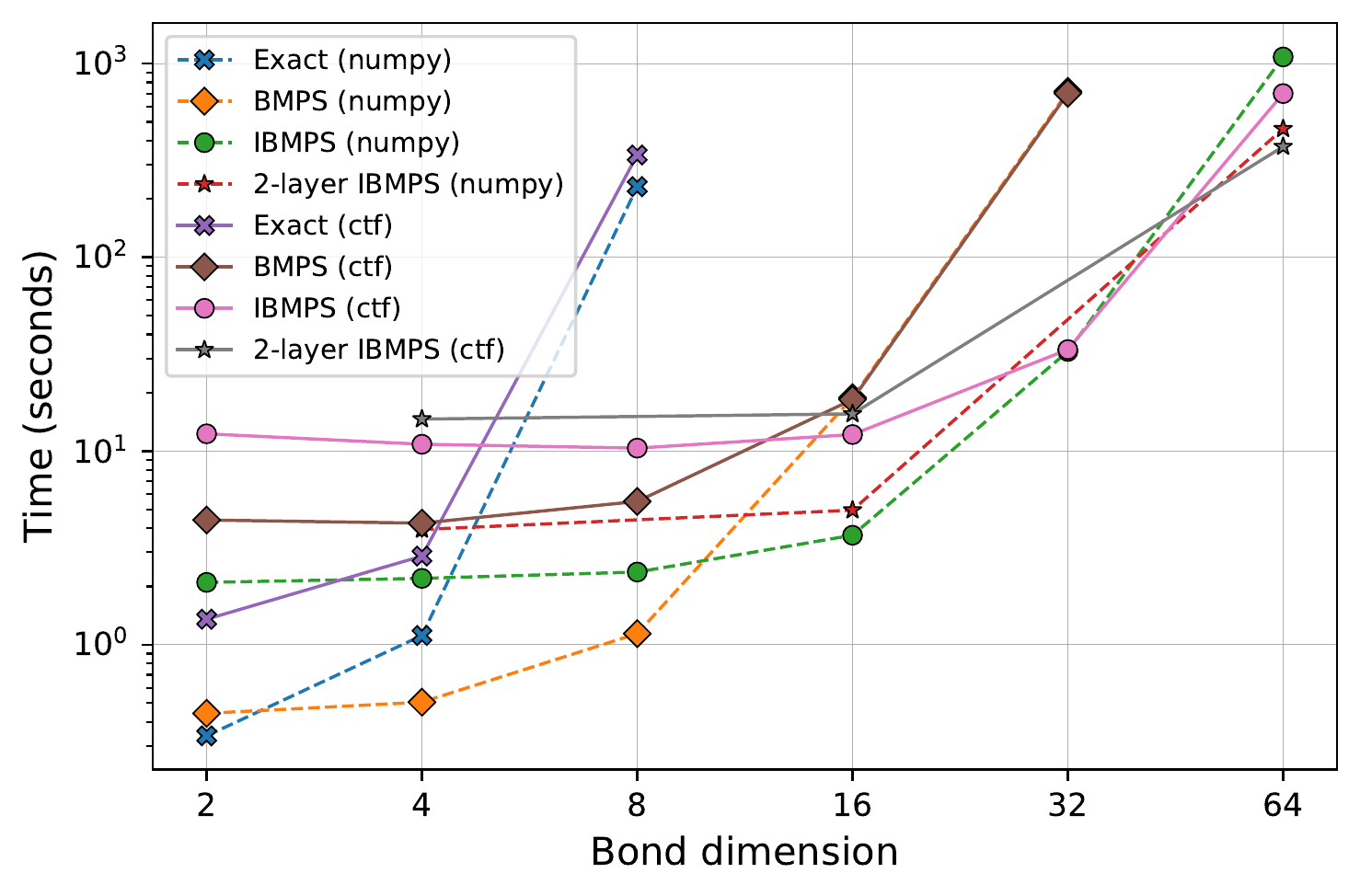}
    \caption{$8 \times 8$ PEPS using $1$ node.}
    \label{fig:contraction-8x8}
\end{subfigure}
\begin{subfigure}[t]{0.95\columnwidth}
    \includegraphics[width=\textwidth]{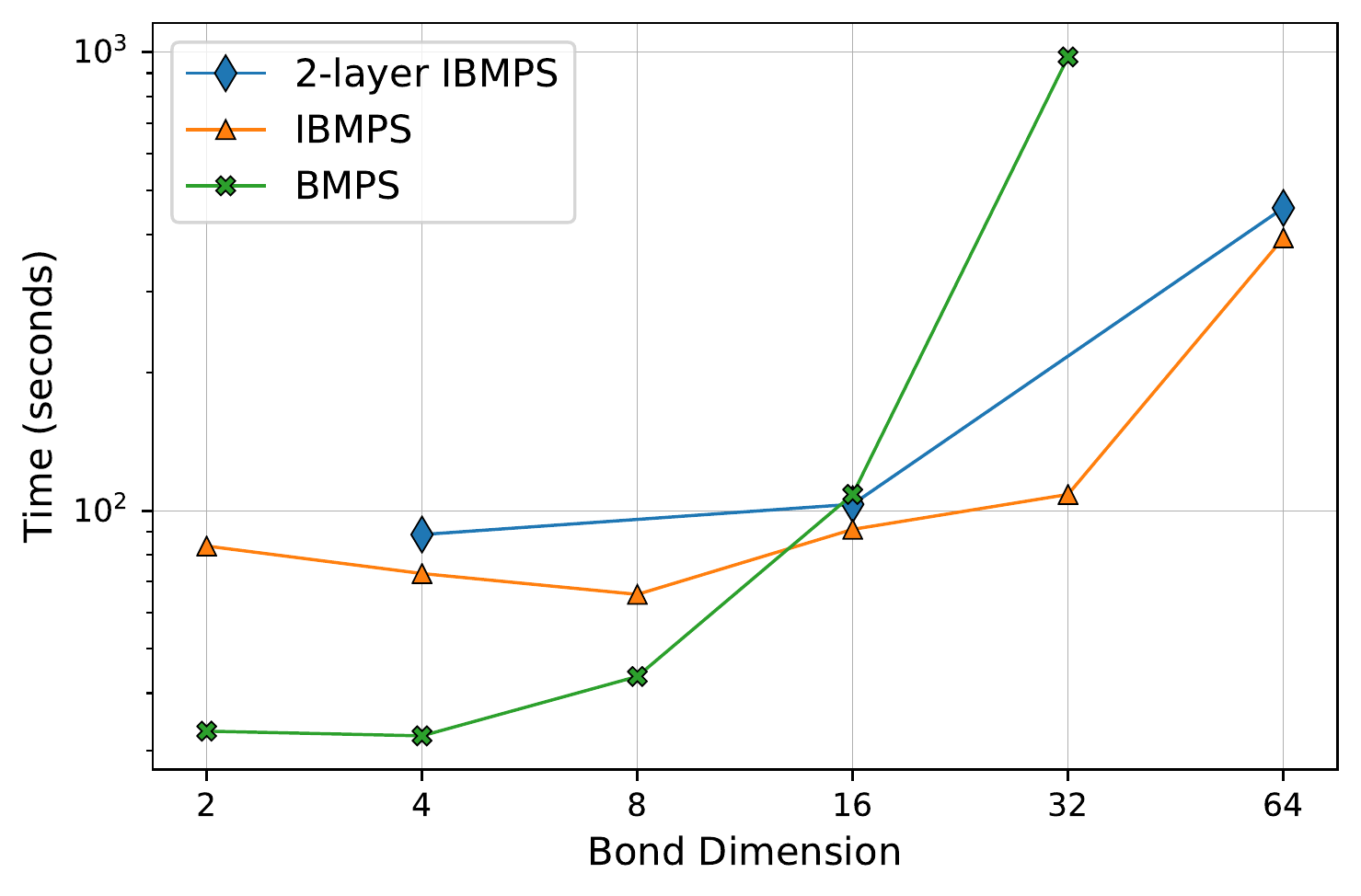}
    \caption{$15 \times 15$ PEPS using $16$ nodes.}
    \label{fig:contraction-15x15}
\end{subfigure}
\caption{Running time for fully contracting a PEPS as the bond dimension grows.}
\label{fig:contraction-time}
\end{figure}

Figure \ref{fig:contraction-8x8} shows the performance of contracting an $8\times 8$ PEPS on a single KNL node of Stampede2 using BMPS, IBMPS, two-layer IBMPS, and an exact algorithm \cite{guo2019simulation} with NumPy and Cyclops as backends. 
Since contracting the inner product between two identical PEPS limits the selection of bond dimensions, we directly generate a PEPS without physical indices to obtain more data points. 
However, two-layer IBMPS is only applicable for the inner product, so we include fewer data points.
%determines it has to be performed on a two-layer PEPS structure, thus leads to fewer data points and additional time for contracting the inner product on the plot. This, however, would not affect our conclusion on its advantage over other methods being compared.

The contraction bond dimension is set equal to the initial bond dimension, which varies from 2 to 64. The trade-off between backends is similar to the PEPS evolution benchmark in that Cyclops is more scalable despite NumPy being faster for small bond dimensions. Further, we show that IBMPS and two-layer IBMPS not only have an advantage on asymptotic computational complexity, but are also more memory-efficient compared to BMPS. In practice, we observe that only IBMPS and two-layer IBMPS are able to contract an $8\times 8$ PEPS with initial and contraction bond dimension both equal to 64 on a single node.
Moreover, we test the highest achievable bond dimension for contracting a $6\times 6$ PEPS using various algorithms on a single node. The exact algorithm and BMPS can only contract such a PEPS with bond dimension less than 30 and 40 respectively, while IBMPS can achieve a bond dimension of 95 and two-layer IBMPS can perform contraction with a bond dimension of more than 100.

We show the performance of contracting a $15\times 15$ PEPS on 16 nodes using Cyclops in Figure~\ref{fig:contraction-15x15}. As in the $8\times 8$ case, we observe that IBMPS has an asymptotic advantage over BMPS, agreeing with our analysis in Table~\ref{tab:ctr_costs}.

%  \begin{figure*}[htbp]
%  \centering
%  \begin{subfigure}[t]{0.49\textwidth}
%      \includegraphics[width=\textwidth]{figures/contract_8x8.pdf}
%      \caption{$8 \times 8$ PEPS with $1$ node}
%      \label{fig:contraction-8x8}
%  \end{subfigure}
%  \begin{subfigure}[t]{0.49\textwidth}
%      \includegraphics[width=\textwidth]{figures/contract_15x15.pdf}
%      \caption{$15 \times 15$ PEPS with $16$ nodes}
%      \label{fig:contraction-15x15}
%  \end{subfigure}
%  \caption{Running time for fully contracting a PEPS versus the rank.}
%  \label{fig:contraction-time}
%  \end{figure*}

\begin{figure}[htbp]
    \centering
    \includegraphics[width=0.95\columnwidth]{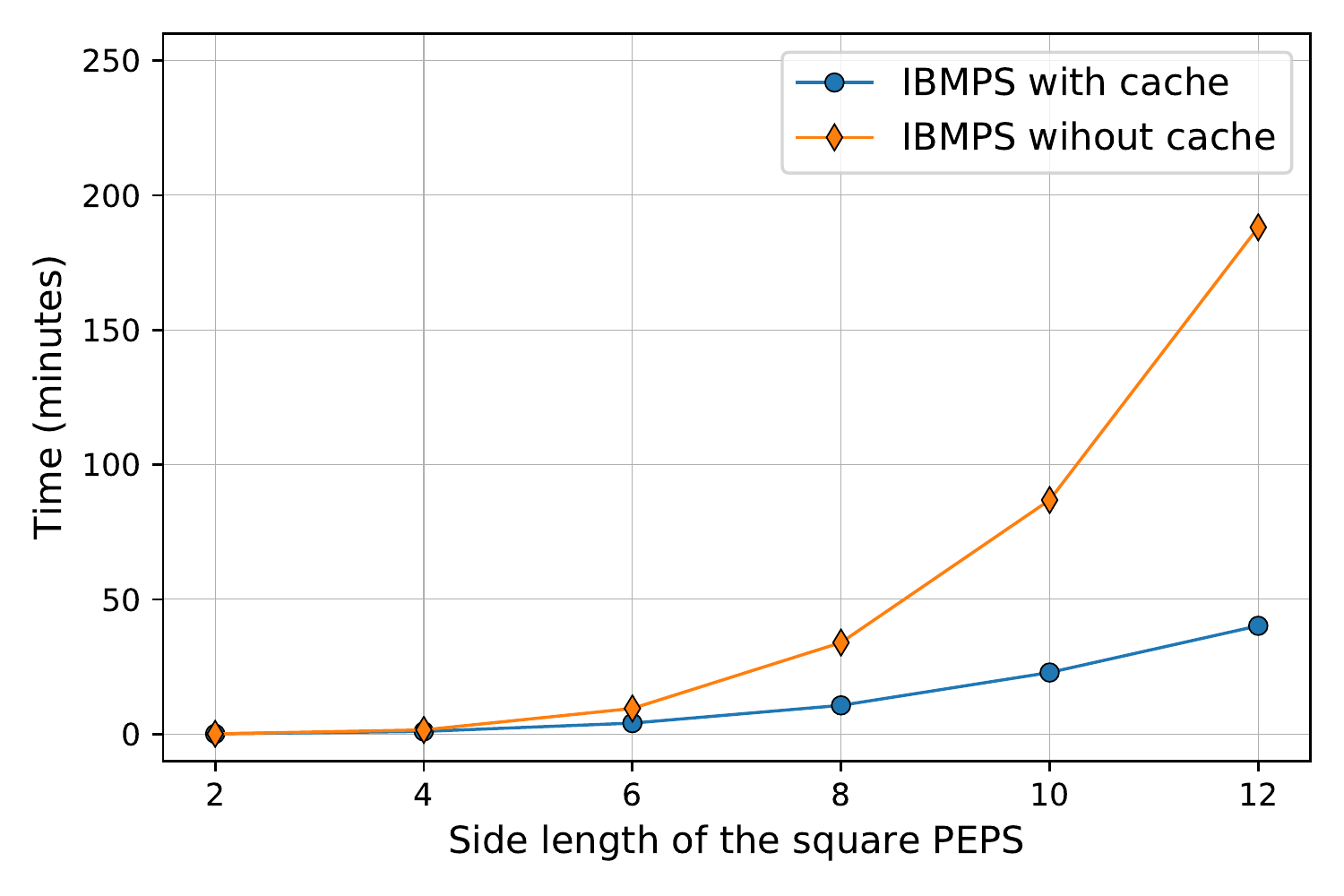}
    \caption{Running time of the expectation value calculation with and without caching for a square PEPS with bond dimension $4$ on 1 node of Stampede2.}
    \label{fig:expectation-cache-time}
\end{figure}

Figure~\ref{fig:expectation-cache-time} studies the performance
of intermediate caching for the expectation value calculation introduced in Section~\ref{subsec:intermediate-caching}.
The expectation operator is composed of one-site operators acting on all sites and two-site operators acting on all pairs of neighboring sites.
With more PEPS sites, the speed-up that caching brings becomes greater. For $12\times 12$ PEPS with bond dimension $4$, the expectation value calculation is 4.7X faster.

\begin{figure}[htbp]
    \centering
    \includegraphics[width=0.95\columnwidth]{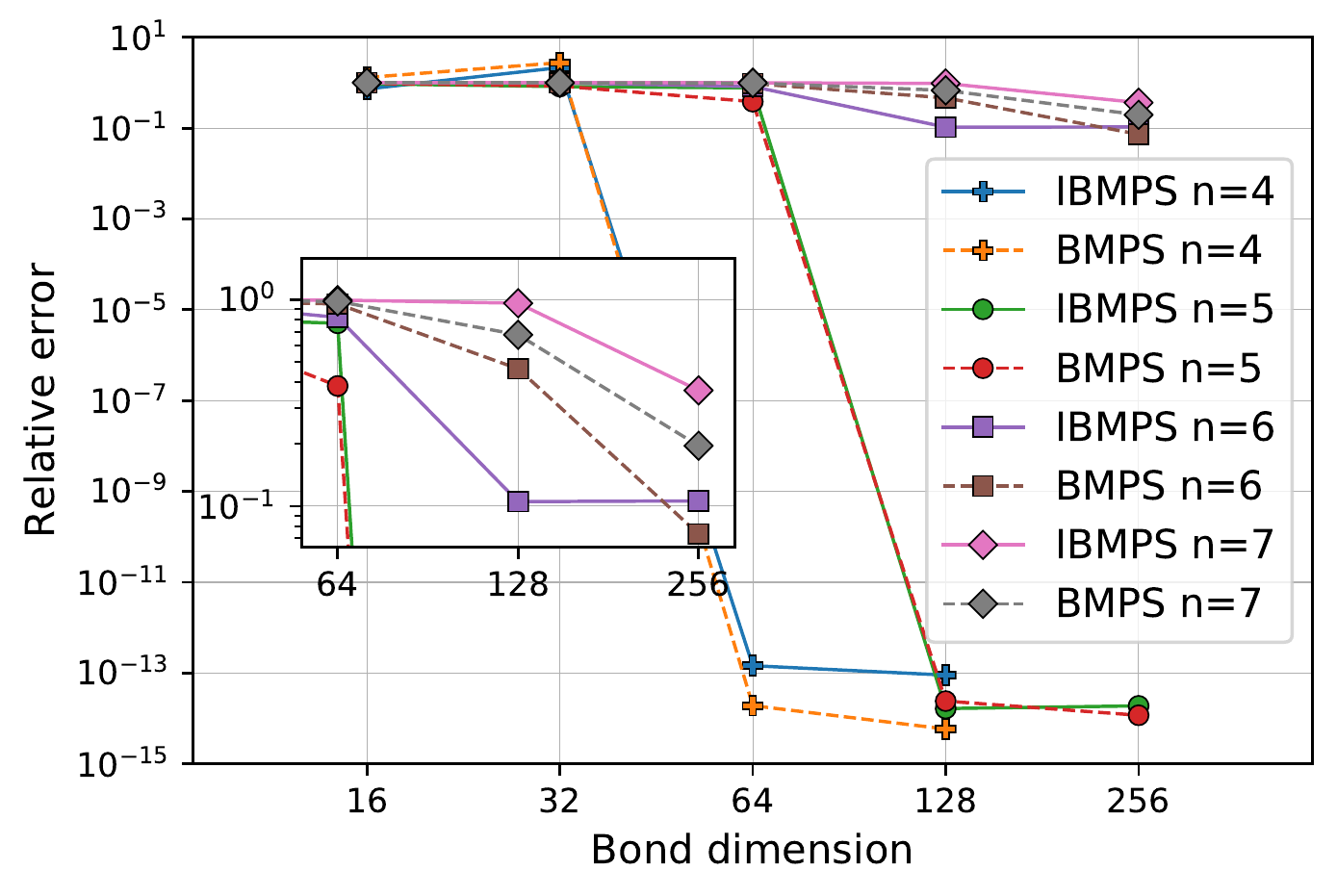}
    \caption{Relative error of contracting an RQC-generated $n\times n$ PEPS of bond dimension 16 using BMPS and IBMPS with varying contraction bond dimension.}
    \label{fig:contraction-rqc}
\end{figure}

We benchmark the accuracy of our contraction algorithms with \emph{random quantum circuits} (RQC) \cite{boixo2018rqc, arute2019rqc}, which are designed to be difficult to simulate classically and have been proven to satisfy both average-case hardness and anti-concentration property \cite{aaronson2017complexity, neill2018blueprint, zhou2020limits, bouland2018quantum}. It is especially challenging to simulate RQCs using tensor networks, since the states created by RQCs are strongly entangled and very sensitive to approximations. By applying our algorithms to this problem, we show that approximate contraction is effective even for one of the hardest simulations.

Following the construction procedure proposed in \cite{arute2019rqc}, we apply iSWAP gates for all pairs of neighboring sites every four layers, increasing the bond dimension by a factor of 4. For the RQC benchmarking, we use $4\times 4$, $5\times 5$, $6\times 6$, and $7\times 7$ PEPS generated by 8 layers of RQC with exact evolution, which has an initial bond dimension of 16. Then, BMPS and IBMPS with varying contraction bond dimension are used to compute one amplitude of the circuit. The result is compared to the exact contraction algorithm to determine the relative error, as shown in Figure \ref{fig:contraction-rqc}. We show that the use of implicit randomized SVD in IBMPS does not incur additional error as compared to the naive SVD. We also observe that after increasing the contraction bond dimension above a certain threshold, the relative error quickly drops to near machine epsilon. This threshold is positively correlated to the PEPS size and the initial bond dimension. For $6\times 6$ and $7\times 7$ PEPS, we achieved one digit of accuracy with a contraction bond dimension of 256 on only 64 nodes with 7808 gigabytes of memory in total.
% \sout{We are limited by the exponential cost of the exact contraction algorithm and the state vector approach to show the relative error of $8\times 8$ and larger PEPS.}

\subsection{Parallel Scaling}

\begin{figure}[htbp]
\centering
    \includegraphics[width=0.95\columnwidth]{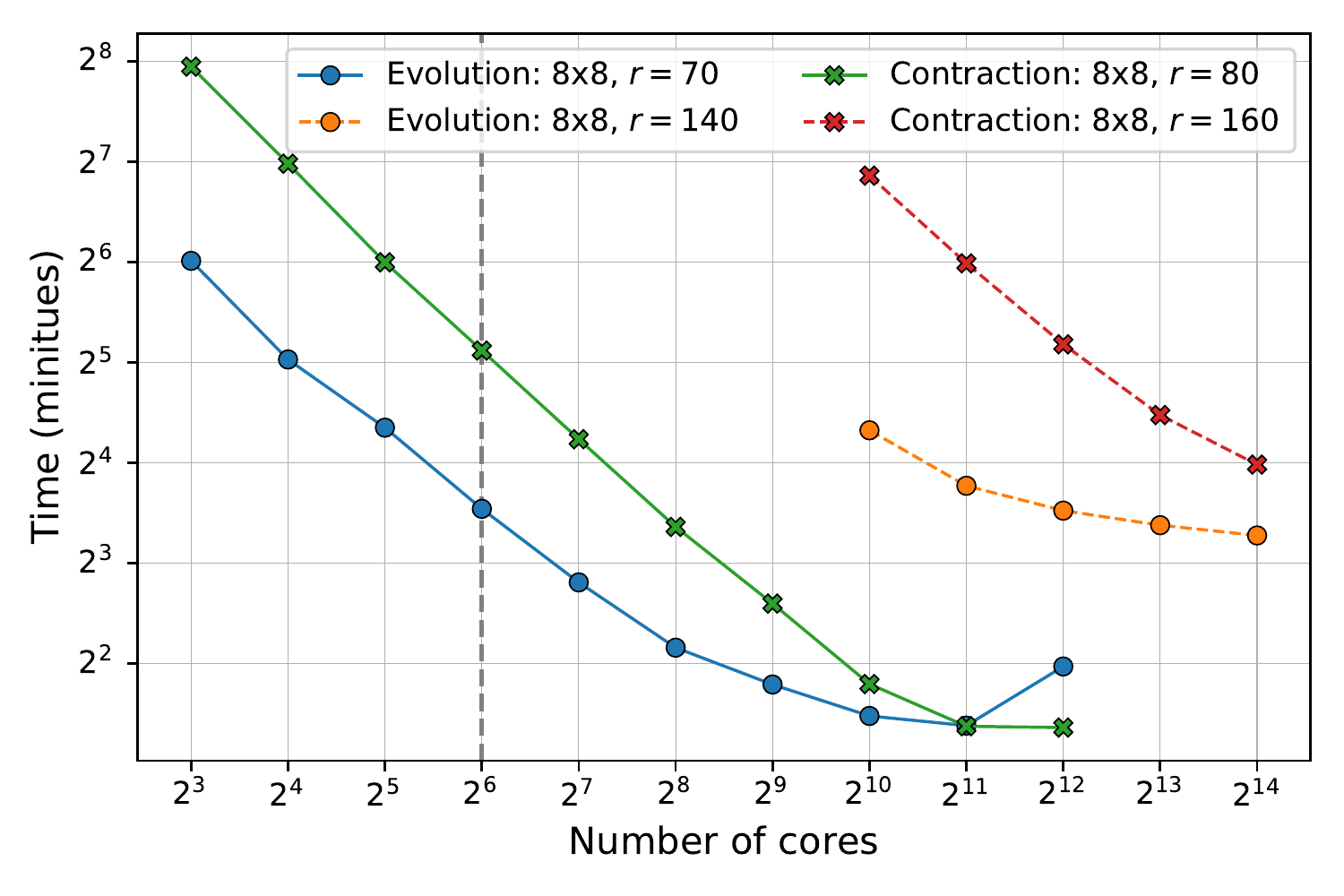}
    \caption{Strong scaling for PEPS evolution (applying one layer of TEBD operators) and PEPS contraction (IBMPS for PEPS with no physical indices); the dashed line separates single-node and multiple-node execution.}
\label{fig:scaling_ss}
\end{figure}

\begin{figure}[htbp]
\centering
    \includegraphics[width=0.95\columnwidth]{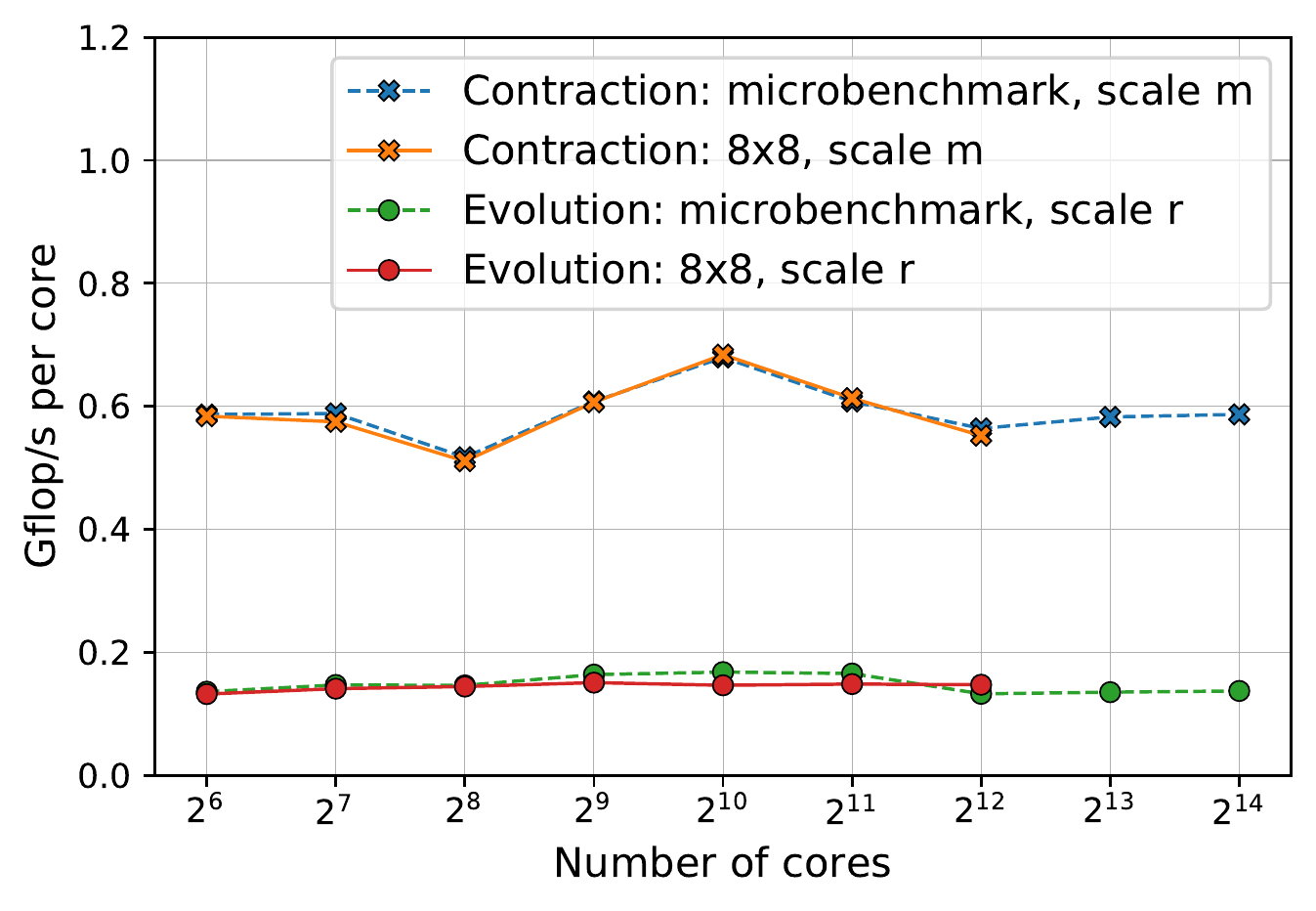}
    \caption{Weak scaling for PEPS evolution (applying one layer of TEBD operators) and PEPS contraction (IBMPS for PEPS with no physical indices)
    with evolution bond dimension $r$ and contraction bond dimension $m$:\\
    $r=70,  83,  98, 117, 140, 166, 197, 235, 280$,\\
    $m=80,  95, 113, 134, 160, 190, 226, 269, 320$.
    }
\label{fig:scaling_ws}
\end{figure}

We study the parallel scaling of applying one layer of TEBD operators (PEPS evolution) and IBMPS (PEPS contraction) in Figure~\ref{fig:scaling_ss} and Figure~\ref{fig:scaling_ws}.
In the strong scaling analysis shown in Figure~\ref{fig:scaling_ss}, we test a smaller problem that occupies most of the memory on a single node and a larger problem that occupies most of the memory on 16 nodes for both PEPS evolution and contraction.
For better performance with Cyclops, we choose to use PPN=16 on 256 nodes ($2^{14}$ cores).
When executing on a single node with multiple cores (less than $2^6$ cores), the running time roughly halves when number of cores doubles, and good strong scaling efficiency is achieved up until all cores of the node are utilized.
%which shows good strong scaling properties when off-node communication cost is small. 

When the benchmarks are executed on the smaller PEPS with multiple nodes, we obtain a speed-up of 4.5X on 32 nodes ($2^{10}$ cores) for PEPS evolution and a speed-up of 13.5X on 64 nodes for PEPS contraction with respect to a single node ($2^6$ cores), after which the performance deteriorates. 
We observe that the fraction time in local matrix multiplication (GEMM), in the smaller PEPS contraction experiments, is 62\% on 1 node, and 16\% on 16 nodes.
For the larger problem starting from 16 nodes, we observe that initially the performance scales well, but for PEPS evolution, the performance deteriorates after 128 nodes, and for PEPS contraction, the performance roughly decreases by one-half on 256 nodes. PEPS evolution has a cost that is only slightly superlinear in the size of the tensor sites, so this kernel tends to be communication-bound.

In the weak scaling analysis shown in Figure~\ref{fig:scaling_ws}, we focus on increasing the bond dimension of PEPS while keeping the memory usage per node constant. This case better reflects the practical usage of large PEPS simulations, where
the major bottleneck is memory and not execution time. As shown in the figure, we observe sustained weak scaling for the full PEPS evolution and contraction benchmark up to 64 nodes ($2^{12}$ cores). Similarly, the time spent in GEMM is always around 60-70\% for PEPS contraction. 
For the experiments with 128 nodes ($2^{13}$ cores) and 256 nodes ($2^{14}$ cores), we instead run microbenchmarks, where only the main computational primitives are benchmarked.
Overall, our approach shows good weak scalability for dealing with large bond dimensions that are prohibitive in cost or infeasible in memory footprint for typical single-node architectures.
% Overall, our approach shows good parallel scalability for high accuracy \rev{and \sout{(}large bond dimension\sout{)}} computations.
% Efficiency falls off gradually if we instead increase the number of sites, since we leverage only the concurrency within operations local to one or two sites.
%in the regime of primary interest (increasing bond dimension to improve accuracy).

\begin{figure*}[htbp]
\centering
\begin{subfigure}{\columnwidth}
    \includegraphics[width=0.95\textwidth]{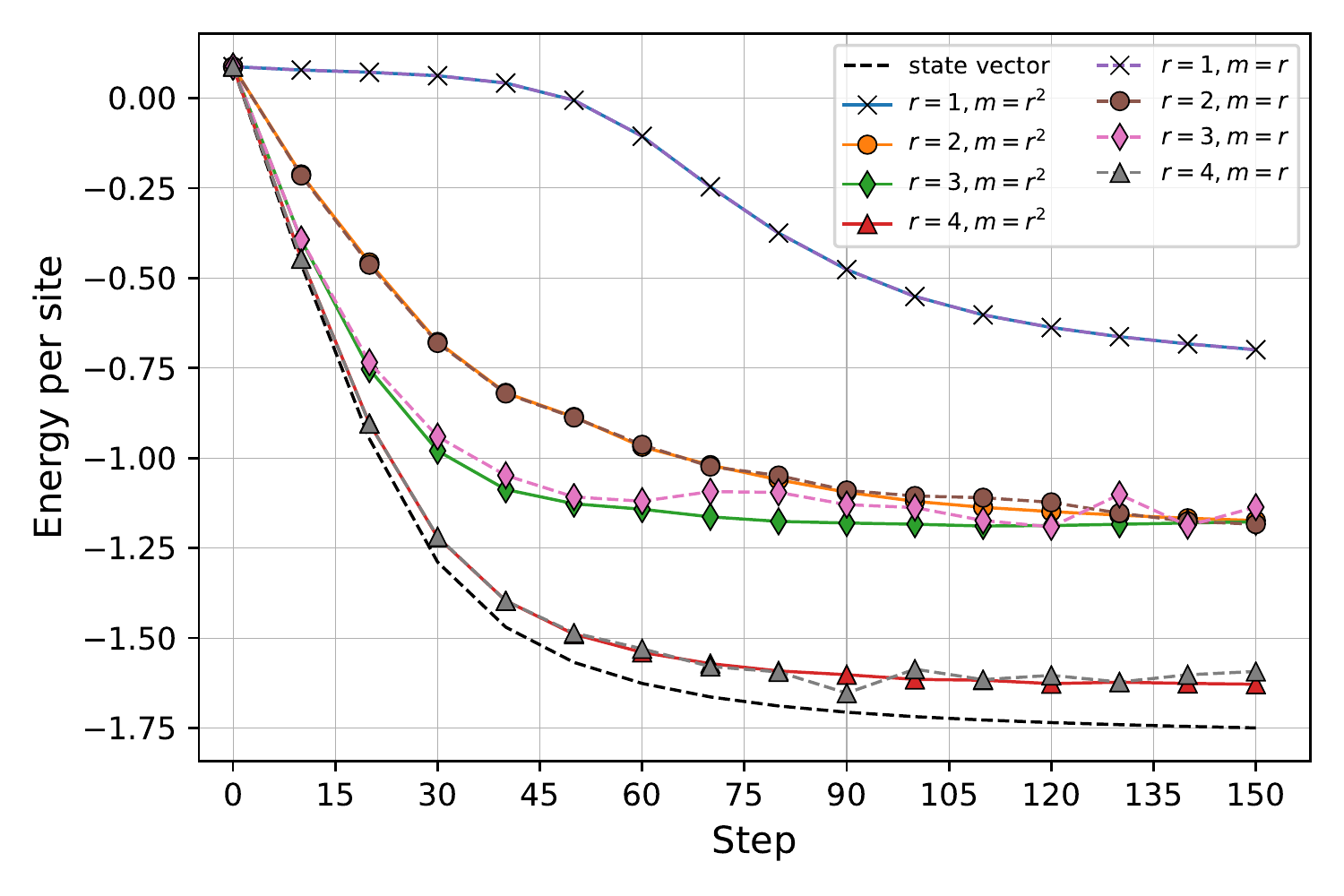}
    \caption{Energies calculated by IBMPS at each step for small bond dimensions.}
\end{subfigure}
\begin{subfigure}{\columnwidth}
    \includegraphics[width=0.95\textwidth]{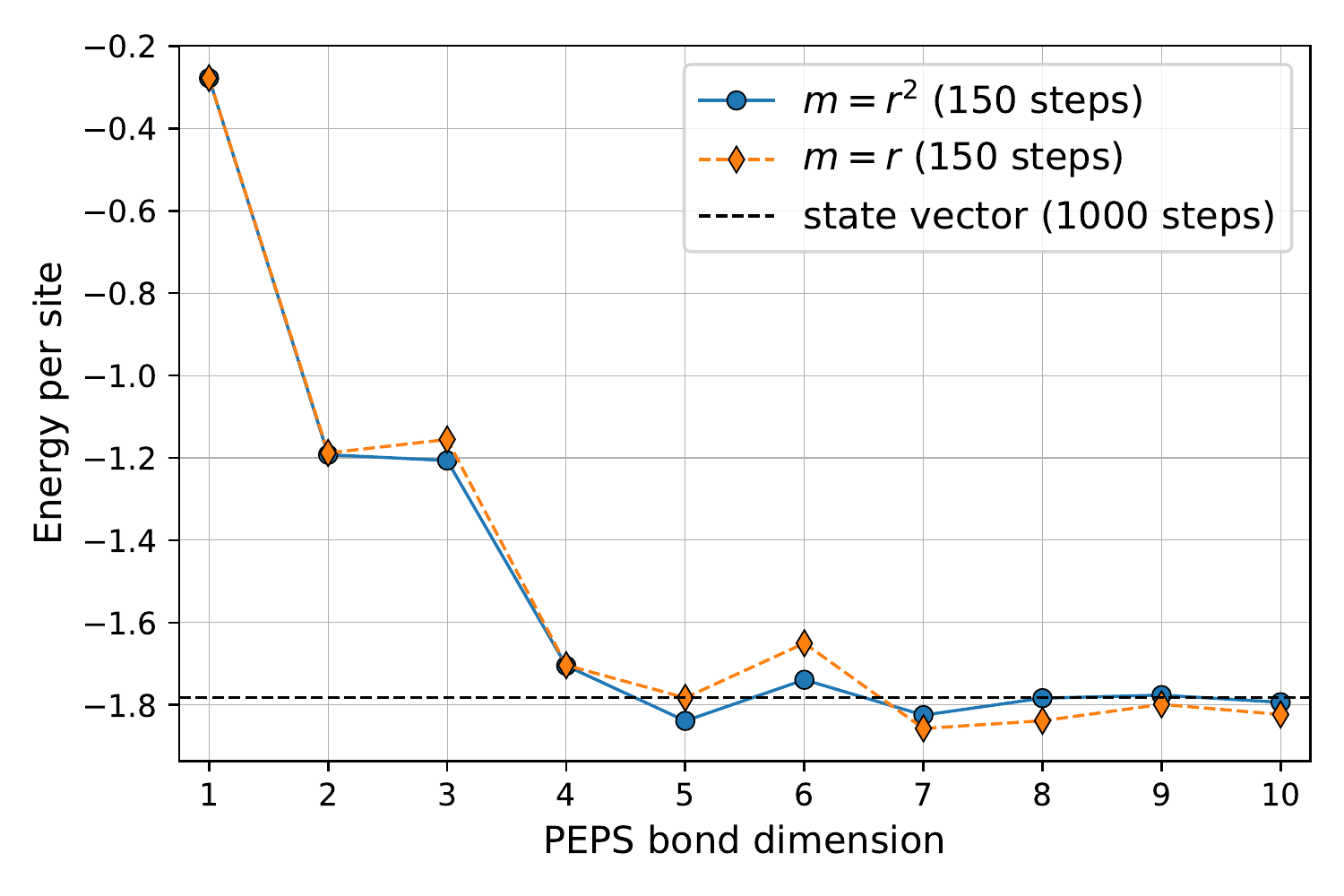}
     \caption{Energies calculated by IBMPS after 150 steps as the bond dimension grows.}
\end{subfigure}
\caption{PEPS ITE results of the $4\times 4$ J1-J2 model with $J_1^x=J_1^y=J_1^z=1.0$, $J_2^x=J_2^y=J_2^z=0.5$ and $h^x=h^y=h^z=0.2$.}
\label{fig:ite-1}
\end{figure*}

% \begin{figure*}[htbp]
%     \centering
%     \begin{subfigure}{\columnwidth}
%         \includegraphics[width=0.95\textwidth]{figures/vqe3x3-snake.pdf}
%         \caption{Convergence for various bond dimensions.}
%     \end{subfigure}
%         \begin{subfigure}{\columnwidth}
%         \includegraphics[width=0.95\textwidth]{figures/vqe3x3-snake-zoom.pdf}
%     \caption{The lower portion of (a) near the ground state energy.}
%     \end{subfigure}
%     \caption{VQE simulation for the $3\times 3$ TFI model with product state initialization.}
%     \label{fig:vqe}
% \end{figure*}

\subsection{Applications}\label{subsec:apps}

We apply the PEPS algorithms to imaginary time
evolution and variational quantum eigensolver simulation to solve for the ground states of quantum systems.

\subsubsection{Imaginary Time Evolution}
We simulate the spin-$\frac{1}{2}$ J1-J2 Heisenberg Model \cite{Yu_2012}
with the Hamiltonian defined as the operator,
\begin{align}\label{eq:j1-j2}
    \mat{H}  =&\:\sum_{\langle i\,j \rangle}
    \left(
        J_1^x \mat{X}_i\mat{X}_j +
        J_1^y \mat{Y}_i\mat{Y}_j +
        J_1^z \mat{Z}_i\mat{Z}_j
    \right) \nonumber\\
     & + \sum_{\llangle i\,j \rrangle}
    \left(
        J_2^x \mat{X}_i\mat{X}_j +
        J_2^y \mat{Y}_i\mat{Y}_j +
        J_2^z \mat{Z}_i\mat{Z}_j
    \right) \nonumber\\
     & +\sum_i \left(h^x \mat{X}_i + h^y \mat{Y}_i + h^z \mat{Z}_i\right),
\end{align}
where $\mat{X}$, $\mat{Y}$, and $\mat{Z}$ are Pauli operators and the indices/subscripts appearing in each Pauli operator indicate the site the operators act on.
%local sites in a 2D lattice that are indexed by subscripts, 
The notation $\sum_{\langle i\,j\rangle}$ denotes a sum over pairs of sites ($i$ and $j$) that are adjacent on a 2D lattice, while $\sum_{\llangle i\,j\rrangle}$ is the same for pairs of sites that are diagonally adjacent (e.g., sites at lattice points $i=(k,l)$ and $j=(k+1,l-1)$).
%iteration over nearest neighbors, i.e., the set $\{(i,j) : |i-j|=1\}$ $\llangle i\,j \rrangle$ represents next nearest neighbors, 
Further, $J_n$ defines the coupling constants between neighbors, and $h$ represents the strength of a transverse magnetic field along a particular axis.

We perform ITE steps as described in Section~\ref{subsubsec:tebd} on PEPS with evolution bond dimension $r$ and contract the resulting PEPS using IBMPS with contraction bond dimension $m$.
Figure~\ref{fig:ite-1} shows the iterations of PEPS ITE for
small bond dimensions and the result of PEPS ITE after 150 steps as the bond dimension grows for a $4\times 4$ J1-J2 model.
The energy of PEPS calculated by IBMPS converges to the ground state calculated by the exact simulation (using the full state vector) after 1000 ITE steps when we increase the bond dimension.
Also, we compare the choice
of contraction bond dimension $m=r^2$ and $m=r$ and observe that for this model, their accuracy is similar while the latter requires much less computation.
% \sout{
% Figure~\ref{fig:ite-1} shows the iterations of PEPS ITE for
% small bond dimensions and the result of PEPS ITE after 150 steps as the bond dimension grows for a $4\times 4$ J1-J2 model with $J_1^x=J_1^y=J_1^z=1.0$, $J_2^x=J_2^y=J_2^z=0.5$ and $h_x=h_y=h_z=0.2$.
% The energy of PEPS calculated by IBMPS converges to the ground state calculated by state vector after 1000 ITE steps when we increase the bond dimension.
% Also, we compare the choice
% of contraction bond dimension $m=r^2$ and $m=r$ and observe that the accuracy of them are similar while the latter requires much less computation.
% }
% Figure~\ref{fig:ite-1} shows the iterations of PEPS ITE for
% small bond dimensions and the result of PEPS ITE after 150 steps as the bond dimension grows for a $4\times 4$ J1-J2 model.
% The energy of PEPS calculated by IBMPS converges to the ground state calculated by state vector after 1000 ITE steps when we increase the bond dimension.
% The same experiments are run with another J1-J2 model that requires larger bond dimensions as shown in Figure~\ref{fig:ite-2}. While the PEPS with $r=10$ cannot converged to the ground state, we still see increasing bond dimension leads to better accuracy, which justifies the need for large bond dimension to simulate complex systems.
% Also, we compare the choice of contraction bond dimension $m=r^2$ and $m=r$ and observe that in either case the accuracy of them is similar while the latter requires much less computation.

\subsubsection{Variational Quantum Eigensolver Simulation}

\begin{figure}[htbp]
    \centering
    \includegraphics[width=0.95\columnwidth]{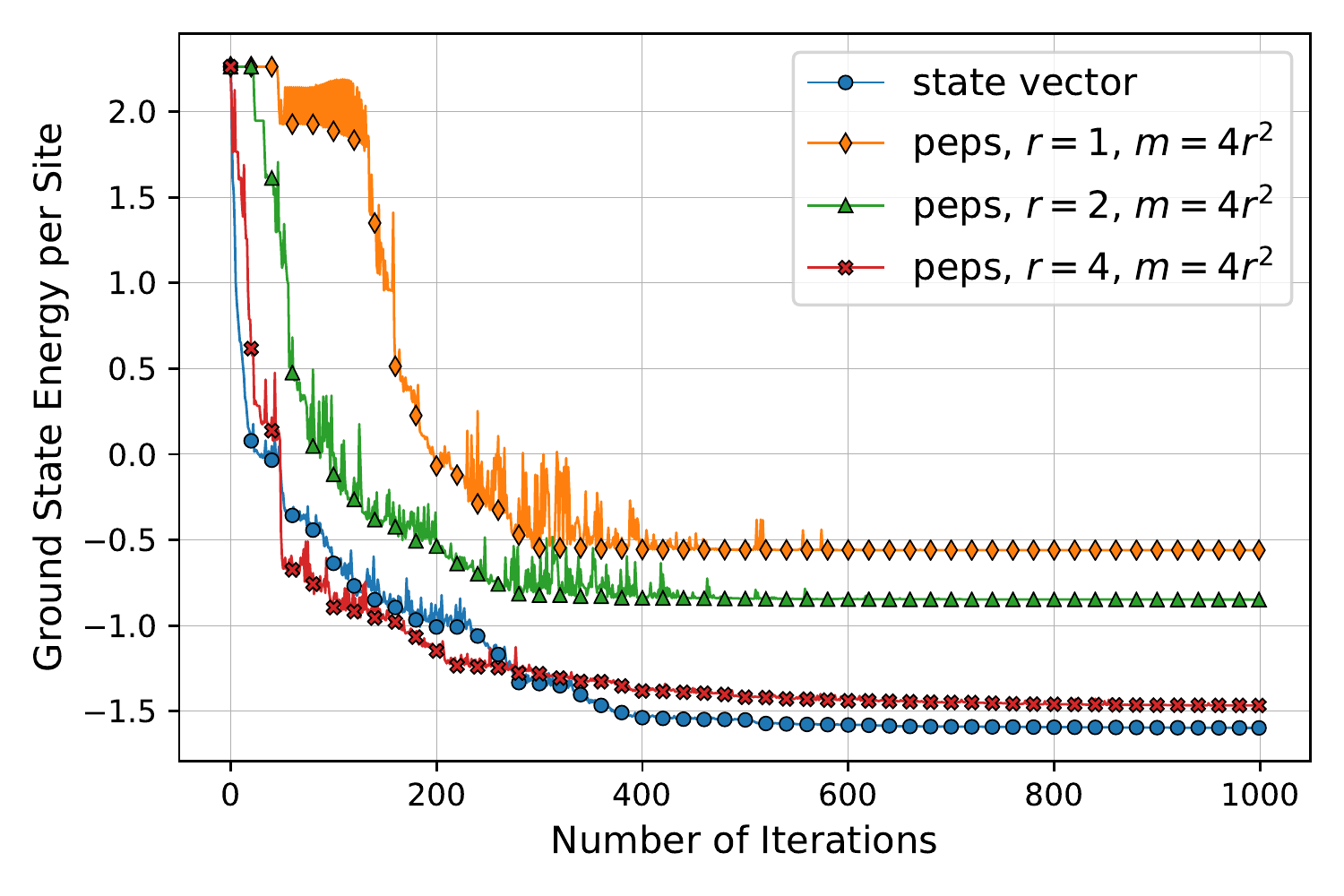}
    \caption{VQE simulation for the $4\times 4$ J1-J2 model with product state initialization. Energies are calculated by IBMPS. Every 20 iterations, markers show the minimum energies found by the simulation up to that point.}
    \label{fig:vqe}
\end{figure}

%We benchmark VQE with the \emph{transverse-field Ising} (TFI) Model with the Hamiltonian defined as
%\begin{equation}\label{eq:tfi}
%\mat{H}=\sum_{\langle i\,j \rangle}
%J^z \mat{Z}_i\mat{Z}_j + \sum_i h^x \mat{X}_i,
%\end{equation}
%which is a special case of Equation \eqref{eq:j1-j2}
%when $J_1^x=J_1^y=J_2^x=J_2^y=J_2^z=h^y=h^z=0$.
%We set $J^z = -1$ and $h^x = -3.5$ in Equation~\eqref{eq:tfi} to simulate the Ferromagnetic TFI Model \cite{motta2019qite}.

We benchmark VQE with the spin-$\frac{1}{2}$ J1-J2 Heisenberg Model (see Equation~\eqref{eq:j1-j2}).  The parameters are set as $J_1^x=J_1^y=J_1^z=1.0$, $J_2^x=J_2^y=J_2^z=0.5$ and $h_x=h_y=h_z=0.2$.

%We utilize the \emph{sequential least squares programming} (SLSQP) optimization algorithm \cite{kraft1988software}, provided via
%the \texttt{scipy.optimize.minimize} function in Python.
We utilize the \emph{constrained optimization by linear approximation} (COBYLA) algorithm \cite{powell1994optimization}, provided via
the \texttt{scipy.optimize.minimize} function in Python.
Our variational ansatz, which is passed to the optimizer, applies a parameterized quantum circuit composed of repeated layers to construct the state prior to calculating the expectation value of the energy.
Each circuit layer consists of single rotation gates $\mat{R}_y(\theta) = e^{-i\theta\mat{Y}/2}$ applied to each qubit. The rotation gates are followed by a set of CNOT gates, which are applied to each nearest neighbor pair.
% A schematic of one layer for a $2 \times 2$ qubit system is shown in Figure~\ref{fig:vqe_single_layer}.

%\sout{With our ansatz, we find that the number of circuit layers needed for good accuracy is on the order of the number of qubits. 
%\edgar{I am skeptical we've scaled sufficiently far out to conclude above...}
%However, as the number of circuit layers is increased by one, the maximum bond dimension of the operator applied to PEPS increases exponentially due to the application of two-qubit gates.  The maximum bond dimension of the PEPS after application of a certain number of circuit layers without truncation is equal to $2^{2n}$, where $n$ is the number of layers.  Since we truncate PEPS down to a maximum bond dimension of 4 or less, we are only able to obtain reasonable results with three or fewer layers.  Therefore, we constrain our algorithm so that three circuit layers are applied to 3x3 systems.}

%
%With our ansatz and three layers, we observe in Figure~\ref{fig:vqe} that with our 3-by-3 qubit state, exact VQE obtains a ground state energy per site of -3.57049.  With eight layers, we are able to measure an energy of -3.59784, and the exact ground state energy is computed to be -3.60024.  Due to truncation, PEPS reaches a lower level of accuracy for VQE and follows a different optimization trajectory.
Each VQE iteration involves the evolution and contraction of PEPS in order to calculate the expectation value, which creates a substantial bottleneck in terms of running time.  
%As mentioned previously, contraction times for a PEPS with nine qubits, rank greater than 3, and boundary MPS contraction can take several minutes. 
%Since each optimizer iteration computes multiple objective function evaluations, VQE itself involves running hundreds or even thousands of function %evaluations prior to convergence.
VQE itself involves running hundreds or even thousands of objective function evaluations prior to convergence.
Consequently, we focus our accuracy experiments on a small number of qubits with bounded iterations.

As can be seen in Figure~\ref{fig:vqe}, for a $4 \times 4$ qubit system increasing the evolution and contraction bond dimensions used in the PEPS simulation improves the lowest energy per site that each system reaches.
The energy obtained is $-0.55990$, $-0.84822$, and $-1.46659$ for a maximum evolution bond dimension of $1$, $2$, and $4$, and a maximum contraction bond dimension of $4$, $16$, and $64$, respectively. By comparison, the state vector system reaches a value of $-1.59796$, while the approximate ground state energy per site (computed from a state vector ITE run) is $-1.77688$.

\section{Related Work} \label{section:related-work}

While algorithms and parallel computations with PEPS are a relatively new area, there exist several works that have taken similar approaches to ours.
The PEPS++ method has been proposed by He et al. as a computational method that enables parallel computation for quantum systems simulation~\cite{he2018peps}. To overcome the high order of complexity, i.e. $O(r^{10})$, they combine quantum Monte Carlo methods with PEPS algorithms to avoid the two-layer PEPS contraction. Instead, they perform one-layer PEPS contraction along with a sampling procedure, which reduces the asymptotic cost to $O(r^{6})$ for each sampling step. In the implementation, they use processes-level parallelism for sampling steps and thread-level parallelism for tensor computations, which differs from the distributed-memory tensor computation considered in this work.

Like our work, Guo et al. have carried out large-scale parallel quantum circuit simulations utilizing PEPS~\cite{guo2019simulation}. They simulate random quantum circuits exactly by applying gate operators on PEPS without truncation, and they calculate amplitudes by projecting the resulting PEPS onto the corresponding basis states. The cost of this method is associated with the entanglement generated by the quantum circuit instead of the circuit size, but could still be exponential because it requires exact PEPS contraction. They also use Cyclops for distributed-memory tensor computations, and their results include an exact contraction of a $7\times 7$ PEPS with bond dimension 32 using 4096 nodes of the Tianhe-2 supercomputer. Our work differs from theirs in that we focus on approximate PEPS algorithms with polynomial time complexity, which allows for larger bond dimensions with a controlled error and consequently serves different application needs.

A related algorithm, the \emph{higher-order tensor renormalization group} (HOTRG) has also been studied in a distributed parallel setting~\cite{yamada2018optimization}, where an optimal reordering procedure for tensor contractions used in HOTRG is proposed. In comparison, our work provides a more general approach via the use of Cyclops, which automates the performance optimization within tensor contractions.
We also consider a broad set of PEPS primitives as opposed to a single component of a PEPS contraction algorithm.

There are also a number of libraries available for quantum system simulation using tensor networks, such as ITensor~\cite{itensor}, NCON~\cite{pfeifer2014ncon}, quimb~\cite{gray2018quimb}, TeNPy~\cite{hauschild2018tenpy}, Uni10~\cite{kao2015uni10}, and qTorch~\cite{fried2018qtorch}. These libraries implement useful algorithms in the field with good sequential, threaded, or GPU performance. Our library, Koala, contributes to this community with the use of distributed-memory parallelism.

% tensor/machine leanring?: TensorLy~\cite{kossaifi2019tensorly},

\section{Conclusion} \label{section:conclusion}

As demonstrated by our application studies, tensor network methods can systematically improve the accuracy for approximate simulation of quantum systems and near-term quantum devices by using larger bond dimensions (i.e. larger tensors). The new algorithms and Koala library we introduced provide an efficient approach for PEPS evolution and contraction and enable the use of distributed-memory tensors. Performance studies on the Stampede2 supercomputer indicate that our approach provides scalable PEPS primitives and systematically improvable accuracy.

\section*{Acknowledgments}
We would like to thank Phillip Helms, Ryan Levy, Garnet Chan, and Bryan Clark for helpful conversations and insights on tensor network methods.
Yuchen Pang, Annika Dugad, and Edgar Solomonik were supported via the US NSF OAC RAISE/TAQS program, award number 1839204.
This work used the Extreme Science and Engineering Discovery Environment (XSEDE), which is supported by National Science Foundation grant number ACI-1548562.
We used XSEDE to employ Stampede2 at the Texas Advanced Computing Center (TACC) through allocation TG-CCR180006.

% \newpage
\bibliographystyle{IEEEtran}
\bibliography{IEEEabrv,references}

\end{document}